\documentclass[letterpaper, 10 pt, conference]{ieeeconf}  

\IEEEoverridecommandlockouts                              

\usepackage{graphicx} 
\usepackage{subfigure}
\usepackage{multirow}
\usepackage{amsmath} 
\usepackage{cite}
\usepackage{siunitx}
\usepackage{booktabs}
\usepackage[T1]{fontenc}
\usepackage[utf8]{inputenc}
\title{\LARGE \bf Blockchain-based Decentralized Co-governance: Innovations and Solutions for Sustainable Crowdfunding}

\author{Bingyou Chen*, Yu Luo*, Jieni Li, Yujian Li, Ying Liu, Fan Yang, Junge Bo and Yanan Qiao** 
\thanks{E-mail(s): chenbill718@gmail.com, \{1637365835, lijieni0622, 750746568, 1281026133\}@qq.com, f.yangcs@gmail.com, bojunge@xjtu.edu.cn}
\thanks{This work was supported by National Training Program of Innovation and Entrepreneurship for Undergraduates.}
\thanks{*Yu Luo and Bingyou Chen contribute equally. Faculty of Electronic and Information Engineering, Xi'an Jiaotong University, Shaanxi, China.}
\thanks{**Corresponding Author: Y. Qiao. qiaoyanan@mail.xjtu.edu.cn}
}

\begin{document}

\maketitle
\thispagestyle{empty}
\pagestyle{empty}

\begin{abstract}

This thesis provides an in-depth exploration of the Decentralized Co-governance Crowdfunding (DCC) Ecosystem, a novel solution addressing prevailing challenges in conventional crowdfunding methods faced by MSMEs and innovative projects. Among the problems it seeks to mitigate are high transaction costs, lack of transparency, fraud, and inefficient resource allocation. Leveraging a comprehensive review of the existing literature on crowdfunding economic activities and blockchain's impact on organizational governance, we propose a transformative socio-economic model based on digital tokens and decentralized co-governance. This ecosystem is marked by a tripartite community structure - the Labor, Capital, and Governance communities - each contributing uniquely to the ecosystem's operation. Our research unfolds the evolution of the DCC ecosystem through distinct phases, offering a novel understanding of socioeconomic dynamics in a decentralized digital world. It also delves into the intricate governance mechanism of the ecosystem, ensuring integrity, fairness, and a balanced distribution of value and wealth.

\end{abstract}

\section{Introduction}

With the advent and rapid proliferation of the internet, online crowdfunding has emerged as a novel means for Micro, Small, and Medium Enterprises (MSMEs) and innovative projects to secure financial backing. This method, characterized by its low cost, minimal contractual obligations, and high efficiency, has proven to be a valuable supplement to conventional financing methods \cite{cosh2009outside}. A notable example of its success is the Chinese film "Monkey King: Hero Is Back," which raised over a million USD through equity crowdfunding in 2015 and subsequently grossed a billion USD at the box office. This case underscores the untapped potential of civilian capital and the promising future of crowdfunding.

However, both conventional centralized and decentralized crowdfunding i.e. Initial Coin Offering (ICO) without governance mechanisms, are not without their flaws. These models are susceptible to credit risk due to information asymmetry. While projects can start raising funds from the public once they have provided relevant information to the crowdfunding platform and passed an audit, the centralized nature of the platform makes it difficult to supervise and prevent fraudulent activities. Similarly, ICO crowdfunding, despite its decentralized nature, also faces these problems due to the lack of a robust governance mechanism\cite{adhami2019initial}.

\textbf{The first problem is the lack of examination of crowdfunding information.} Crowdfunding projects often struggle to balance transparency and privacy, leading to either the risk of idea theft, as seen with the Pressy project on Kickstarter \cite{doh2010does}, or information opacity that deters funders. This issue is also prevalent in ICOs, with cases like Centra Tech and Telegram ICOs demonstrating fraudulent practices and regulatory intervention. Conventional crowdfunding platforms struggle to provide a trustworthy and effective solution due to potential collusion and lax regulation \cite{grossman1981informational, belleflamme2015economics}.

\textbf{The second problem is the lack of supervision over the raised funds.} Centralized crowdfunding platforms often lack real-time supervision over raised funds\cite{dana2011long, bouvard2018two}, leading to misuse of funds and fraud. This was evident in the case of iBackPack of Texas, LLC. In addition, since the project is in the creative period, the members have a subjective grasp of the execution plan, which leads to the deviation of the specific plan\cite{tucker2011does}. This was evident in the case of ZANO drone project on Kickstarter. Similarly, in the ICO space, the Centra Tech ICO saw raised funds allegedly spent on personal expenses rather than the promised financial products \cite{sec}.

\textbf{The last problem is the lack of sustainable development of the crowdfunding ecosystem.} Many potential funders are deterred by the risks associated with crowdfunding. This tripartite game between fundraisers, funders, and platform governors often ends in a non-cooperative Nash equilibrium due to information asymmetry, limiting participation \cite{kim2022risk, belleflamme2015economics, nash1950equilibrium}. The ICO ecosystem also suffers from sustainability issues, with a study finding that 80 percent of ICOs were scams and only 8 percent reached the trading stage. Thus, there is a need for a new mechanism that promotes cooperation and benefit sharing to boost participation and market growth.

In order to address the above problems, we have proposed a novel approach that leverages blockchain technology to create a decentralized co-governance crowdfunding (DCC) ecosystem. This paper will present several key innovations that address the issues inherent in conventional crowdfunding platforms, which are as follows:

(1) We will introduce the DCC ecosystem, its communities, and the Delegated Proof of Labor Representative (DPoLR) election consensus. This consensus mechanism forms the core of our proposed system, ensuring equitable representation and decision-making within the ecosystem.

(2) We will delve into the tripartite governance architecture that is designed to facilitate benefit-sharing among all participants. This innovative approach transforms the conventional crowdfunding ecosystem governance mechanism, fostering a cooperative rather than competitive environment.

(3) We will present a suite of risk mitigation strategies tailored to DCC activities. These include blockchain-based identity authentication and regulatory compliance, on-chain protection mechanisms for smart contracts, and a decentralized autonomous council-centered architecture.

(4) We will introduce a distributed token evaluation system for concept validation and a distributed, token-incentivized supervision and credit scoring system. Both of these systems are designed to enhance transparency and trust within the ecosystem.

(5) We will provide an economic analysis of the sustainability of the DCC ecosystem. This includes an exploration of game behavior under decentralized co-governance, an examination of how the blockchain consensus mechanism catalyzes crowdfunding enthusiasm through benefit-sharing, and a model of crowdfunding token circulation and valuation.

This thesis is organized as follows: Section 2 reviews the related work such as Steem Blockchain, DAICO, and CGS. Section 3 outlines the various methods used in our research, including literature review, case study analysis, policy and legal analysis, and quantitative analysis. Section 4 introduces our models for the DCC ecosystem communities and the governance architecture. Section 5 discusses strategies for mitigating risks in DCC activities, including blockchain-based identity authentication and on-chain protection mechanisms. Section 6 conducts an economic analysis of the sustainability of the DCC ecosystem, including game behavior, blockchain consensus mechanism, and an economic model of crowdfunding token circulation. Finally, Section 7 summarizes the research contributions of this thesis, discusses the implications for research and practice, and suggests directions for future research.

\section{Related Work}

\subsection{Steem Blockchain Ecosystem}

\begin{figure}[htbp]
	\centering
	\includegraphics[width=1\linewidth]{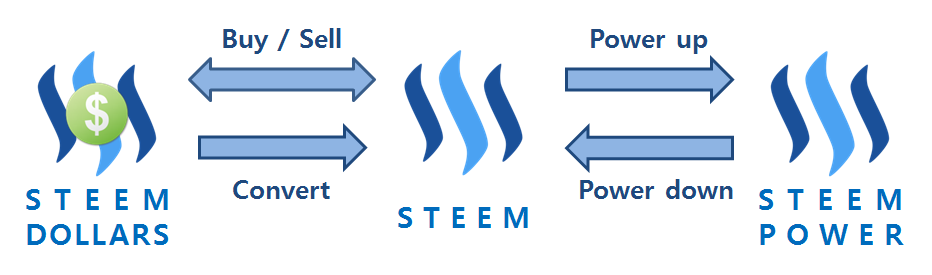}
	\caption{Steem Token Convertion Mechanism}\label{fig2-1}
\end{figure}

The Steem blockchain (See Figure \ref{fig2-1}) has been the subject of various studies due to its unique integration of social networking and blockchain technology. Guidi et al. \cite{guidi2020, guidi2021} conducted comprehensive analyses of the Steem blockchain, focusing on the structure of the transaction graph and the role of witnesses in the Steem blockchain. Their studies revealed important insights into the social and economic aspects of the Steem blockchain, including the detection of bots offering paid services and the social impact of witnesses.

In addition, Cai et al. \cite{cai2018} provided an overview of decentralized applications (dApps) and discussed the future value of blockchain, with Steem being one of the key examples. Their study highlighted the importance of dApps in the development of blockchain systems and the potential of blockchain technology in empowering software systems.

Kim and Chung \cite{kim2018} proposed a process for building a desirable model of a token economy based on the case of Steemit, a blogging and social networking website built on the Steem blockchain. Their study provided valuable guidelines for designing a token economy model, emphasizing the importance of aligning the token model with the business model and establishing strategies for raising token value and managing the token economy system.

Lastly, Pereira et al. \cite{pereira2019} compared blockchain-based platforms and centralized platforms from a managerial perspective. Their study provided insights into the conditions under which blockchain-based platforms, such as Steem, are more advantageous than centralized platforms. They considered factors such as transaction cost, cost of technology, and community involvement in their analysis.

These studies provide valuable insights into the Steem blockchain and its applications, laying a solid foundation for our proposed decentralized co-governance crowdfunding system.

\subsection{Decentralized Autonomous Initial Coin Offerings (DAICO)}

In the current landscape, the ICO ecosystem faces significant governance challenges due to the absence of standard regulation. The power dynamics are skewed heavily in favor of the ICO proponents, such as entrepreneurs and project insiders, leaving token holders, who are the primary contributors of funding, at a disadvantage. Upon the successful completion of an ICO, the entrepreneurs gain complete control over the raised funds, creating a power imbalance.

\begin{figure}[htbp]
	\centering
	\includegraphics[width=1\linewidth]{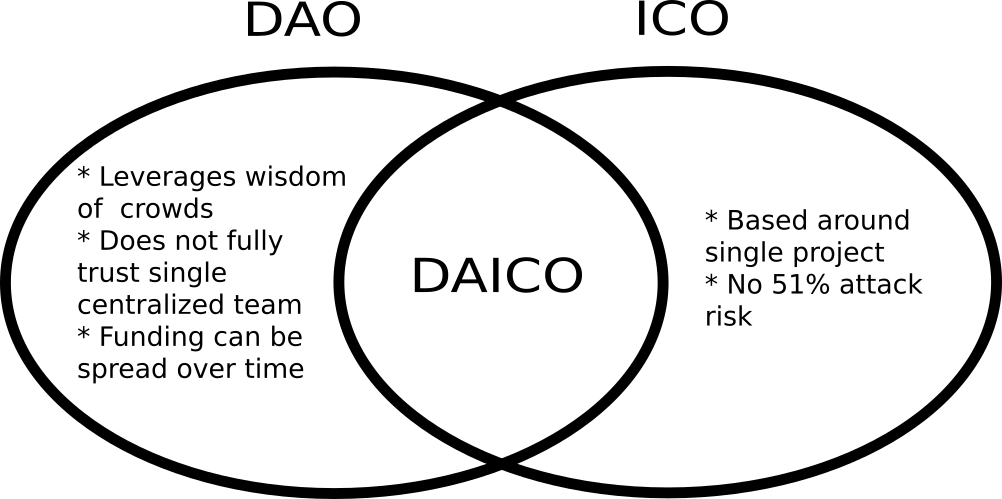}
	\caption{Decentralized Autonomous Initial Coin Offerings Mechanism}\label{fig2-2}
\end{figure}

This governance issue has led to an urgent need for improvements to protect token buyers from scams and mismanagement of funds. While the market awaits a regulatory response, several solutions have been proposed to address these agency problems. One such solution is the Decentralized Autonomous Initial Coin Offering (DAICO) (See Figure \ref{fig2-2}), a mechanism that combines the benefits of Decentralized Autonomous Organizations (DAOs)\cite{buterin2014next} with the ICO fundraising model\cite{buterin2018}.

Unlike traditional ICO tokens, where the smart contract only regulates the inflow of funds in exchange for tokens, the DAICO contract operates in a 'tap mode'. This mode is akin to 'stage financing' in venture capital, where the funds are held in an escrow account and released in tranches according to the milestones set in the initial business plan. If the contributors disapprove of the management conduct or are dissatisfied with the project performance, they can vote to shut down the DAICO and reclaim their cryptocurrencies, leading to the self-destruction of the contract.

DAICOs empower investors by allowing them to exert strong control over the managers. They can autonomously shut down the project at any time, thereby compelling the agent-managers to continuously engage with contributors and adhere to the roadmap proposed at the ICO's inception or seek approval for new plans.

\subsection{Coin Governance System (CGS)}

The Coin Governance System (CGS) is an innovative solution designed to address the moral hazard problem that often arises post-ICO\cite{potop2019coin}. Registered in Spain, the CGS aims to safeguard ICO investors while serving as a managerial tool for ICO companies.

\begin{figure}[htbp]
	\centering
	\includegraphics[width=1\linewidth]{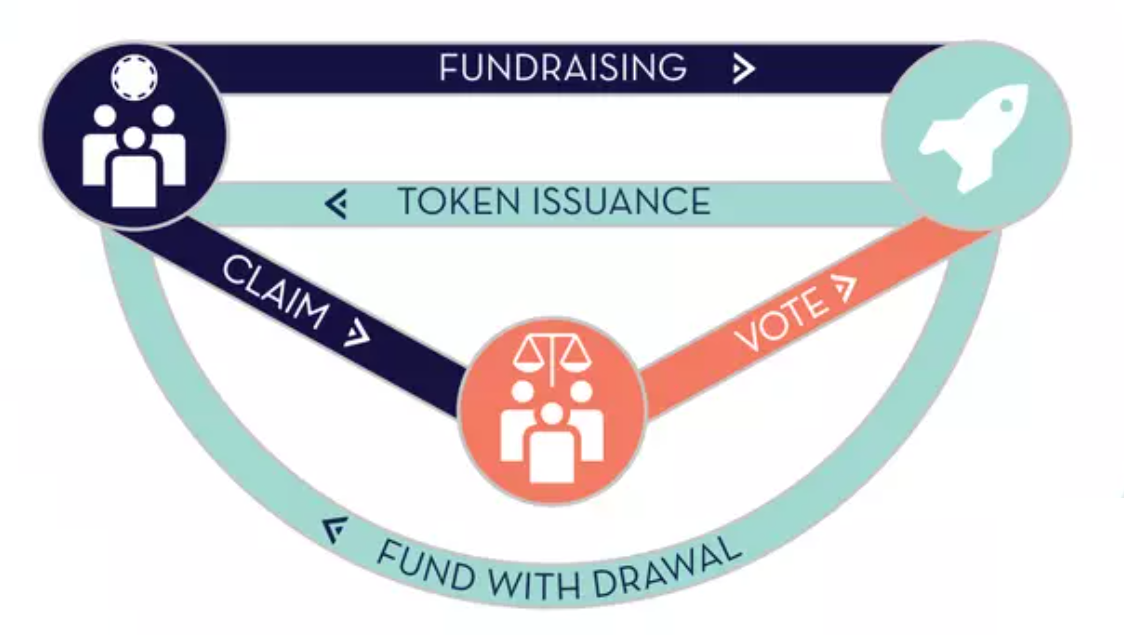}
	\caption{Coin Governance System Structure}\label{fig2-3}
\end{figure}

Upon the completion of a CGS-assisted ICO, the raised capital is held in escrow, similar to the DAICO model. As shown in Figure \ref{fig2-3} The CGS then gradually releases the funds to the ICO initiator. If a token holder suspects that the project is not being executed as promised, they can submit a claim to the CGS by depositing a certain number of tokens. If the claim garners sufficient support from other investors, it is reviewed by a decentralized community of 'CGS Arbiters' who vote on the claim.

The voting process can result in two outcomes: "OK" or "KO". An "OK" outcome means the project continues as planned, while a "KO" outcome triggers a "withdrawal mode" in the CGS smart contract, enabling ICO token holders to withdraw the remaining funds by returning their tokens.

The CGS Arbiters function as a decentralized judicial system, incentivized to vote correctly by earning CGS tokens. Votes are confidential, and arbiters stake a number of CGS tokens in the hope of earning more if their vote aligns with the majority. This mechanism, known as the "prisoner's dilemma", is designed to oversee the behavior of ICO promoters and is a testament to their goodwill to proceed with the ICO.

However, the CGS system introduces a cost for the new venture and adds decision-making uncertainty for entrepreneurs who rely on majority voting to resolve disputes and potentially shut down the business. This system is susceptible to manipulation through online trends and rumors.

In essence, the CGS system is a DAICO contract that elects online participants in its network as arbiters, adding a further level of intermediation between the principal (the token holders) and the agents (the entrepreneurs). While this aligns with the crowd-consensus philosophy of the blockchain to build trust in the community, it may not necessarily act in the best interest of specific ICO investors. Meanwhile, with a general DAICO contract, ICO promoters could increase their token retention rate to maintain a majority and avoid any chance of self-destruction vote by crowd-investors. In such a scenario, the election of external parties as collective judges for issues raised by a minority of unsatisfied token holders can be very impactful in terms of protecting the actual providers of funds.

\section{Research Methodology}

Our research employs a comprehensive approach that combines literature review and case study analysis to explore the challenges and potential solutions in the field of crowdfunding. Then, we will delve into legal analysis and quantitative analysis. These methods will further enrich our understanding of the dynamics of the crowdfunding ecosystem and the potential impact of different governance mechanisms.

\subsection{Literature Review and Case Study Analysis}

We begin our research with an extensive review of academic papers, industry reports, and case studies related to both traditional and ICO crowdfunding platforms. We pay particular attention to the CGS, DAICO, and the Steem Blockchain Ecosystem, aiming to understand their unique features, benefits, and potential challenges.

In parallel with the literature review, we conduct a detailed analysis of various case studies. These cases are selected to represent a wide range of scenarios and outcomes in both traditional and ICO crowdfunding platforms. We also include cases where the CGS, DAICO, and the Steem Blockchain Ecosystem have been implemented.

The case study analysis allows us to understand the practical implications of the issues identified in the literature review. We analyze these cases to identify common patterns, challenges, and potential solutions. This combined approach of literature review and case study analysis provides a robust foundation for our research, enabling us to gain both theoretical and practical insights into the field of crowdfunding.

\subsection{Policy and Legal Analysis}

This subsection provides an analysis of the legal and policy landscape surrounding various forms of crowdfunding in different jurisdictions, with a particular focus on Chinese Mainland, the United States, European countries, and Hong Kong SAR China.

\textbf{Equity-Based Crowdfunding}

In Chinese Mainland, equity-based crowdfunding has not yet gained legal status. Despite early signs of support for the development of private equity-based crowdfunding, recent regulations have tightened, and there is currently no relevant exemption policy for it. In contrast, the United States has embraced equity-based crowdfunding through the Jumpstart Our Business Start-ups Act (JOBS Act), which provides a limited exemption for crowdfunding platforms known as funding portals. Italy, too, has passed legislation similar to the JOBS Act, allowing only innovative start-ups to apply. In Hong Kong, the Securities and Futures Commission (SFC) has issued a consultation paper on the regulation of crowdfunding platforms, indicating a move towards establishing a regulatory framework. However, as of now, there are no specific regulations governing equity-based crowdfunding.

\textbf{Lending-Based Crowdfunding}

Chinese Mainland lacks specific legislation for lending-based crowdfunding, which is broadly regulated under the Contract Part of the Civil Code. In the United States, lending-based crowdfunding is considered an investment contract and is therefore regulated by the Securities Act of 1933 and the Securities Exchange Act of 1934. France, on the other hand, has enacted the Participatory Financing Act, which provides specific regulations for lending-based crowdfunding. In Hong Kong, the Hong Kong Monetary Authority (HKMA) has established a Fintech Supervisory Sandbox, which allows banks and their partnering tech firms to conduct pilot trials of their fintech initiatives. This could potentially cover lending-based crowdfunding platforms, although specific regulations are yet to be established.

\textbf{Reward-Based Crowdfunding}

In Chinese Mainland, reward-based crowdfunding is considered a type of contract and is therefore regulated under the Civil Code. Platforms are required to make clear the qualifications of issuers to avoid illegal fundraising and fraud. In the United States, platforms like Kickstarter avoid legal liability by excluding equity-based crowdfunding and ICO crowdfunding and by clearly stating the rights and obligations of their investors. Hong Kong does not have specific regulations for reward-based crowdfunding. However, the SFC's consultation paper on the regulation of crowdfunding platforms could potentially cover this area in the future.

\textbf{ICO Crowdfunding}

ICO crowdfunding is illegal in Chinese Mainland, with the government banning ICOs entirely due to considerations of stability and real risks. In the United States, ICO crowdfunding is considered an investment contract and is therefore regulated under the Securities Act of 1933 and the Securities Exchange Act of 1934. European countries do not have unified regulations on ICOs, with different countries adopting different approaches. In Hong Kong, the SFC has established a regulatory framework for virtual asset portfolio managers, fund distributors, and trading platform operators. This framework could potentially cover ICO crowdfunding platforms.

Table \ref{table3-0} provides a quick reference for the policy and legal situation for each type of crowdfunding in Chinese Mainland, the USA, Europe, and Hong Kong SAR. In conclusion, the legal and policy landscape surrounding crowdfunding varies significantly across different jurisdictions and types of crowdfunding. As the crowdfunding sector continues to evolve, it is expected that regulatory frameworks will continue to be developed and refined to balance the need for investor protection with the promotion of innovation and growth in this area.

\begin{table*}[ht]
\centering
\resizebox{\linewidth}{!}{%
\begin{tabular}{cccccc}
\toprule
 & \textbf{Chinese Mainland} & \textbf{USA} & \textbf{Europe} & \textbf{Hong Kong SAR} \\
\midrule
\textbf{Equity-Based} & No legal status & JOBS Act & Varies by country & Consultation stage \\
\textbf{Lending-Based} & Regulated by Civil Code & Securities Act & Participatory Financing Act & Fintech Supervisory Sandbox \\
\textbf{Reward-Based} & Regulated by Civil Code & Exclusions in platform agreements & N/A & Consultation stage \\
\textbf{ICO} & Illegal & Securities Act & Varies by country & Regulated for virtual asset managers \\
\bottomrule
\end{tabular}
}
\caption{Policy and Legal Situation for Crowdfunding in Different Jurisdictions}
\label{table3-0}
\end{table*}

\subsection{Quantitative Analysis}

In addition to the policy and legal analysis, our research also involves a quantitative analysis of crowdfunding platforms and projects. This analysis aims to provide a deeper understanding of the dynamics and performance of both conventional centralized crowdfunding platforms and decentralized crowdfunding projects, such as Initial Coin Offerings (ICOs)\cite{shneor2020crowdfunding}. We also conduct an interest analysis of the various stakeholders involved in crowdfunding to understand their motivations, behaviors, and the potential impacts of different crowdfunding models on their interests. This comprehensive quantitative analysis provides valuable insights into the operational mechanisms, performance outcomes, and stakeholder dynamics of different types of crowdfunding.

\subsubsection{Conventional Centralized Crowdfunding Platform}

Since 2016, the sharing economy has been recognized as a significant development strategy by the Chinese government, leading to the emergence of conventional crowdfunding platforms in the country \cite{choi2020sustainable}. However, due to the government's ambiguous stance towards blockchain decentralized economic activities and the stringent crackdown on ICO crowdfunding \cite{okorie2020did}, the crowdfunding economy in China is primarily driven by conventional platforms, positioning China as a major player in conventional commercial crowdfunding activities.

Table \ref{table3-1} provides a brief comparison of the main conventional crowdfunding platforms in China. As Meynhardt et al. \cite{meynhardt2016systemic} pointed out, value co-creation and evaluation services are two key advantages of crowdfunding platforms. Therefore, our analysis primarily focuses on four aspects: value co-creation, strategic services, full link services, and precision financing.

\begin{table*}[htbp]
\centering
\includegraphics[width=1\linewidth]{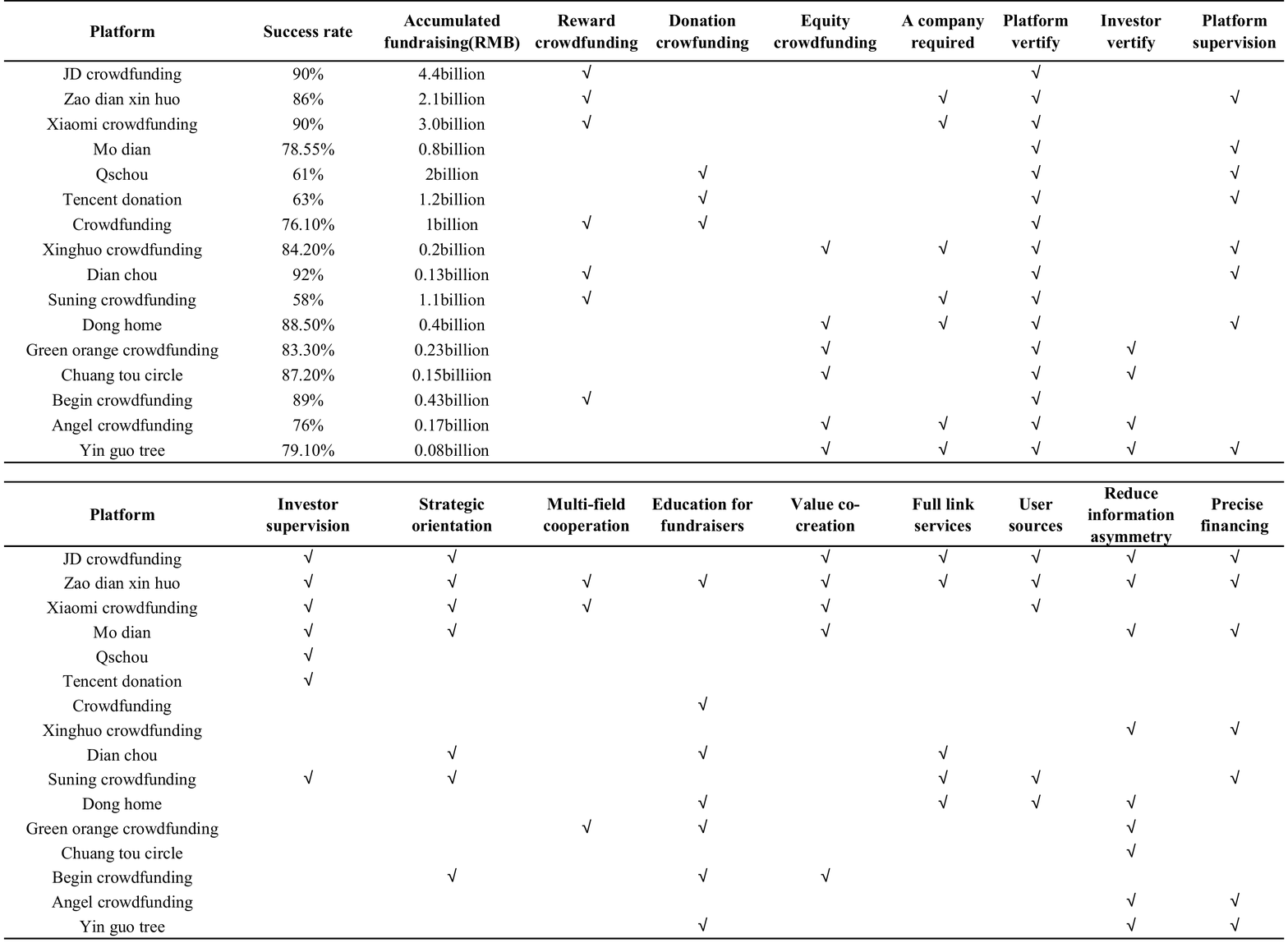}
\caption{Conventional centralized crowdfunding platforms}\label{table3-1}
\end{table*}

According to our survey, reward crowdfunding is the predominant type of crowdfunding in China. Most reward crowdfunding platforms provide a space for investors and fundraisers to exchange ideas about a product, facilitating the co-creation of value. Additionally, most of these platforms analyze product support data and charge for value-added services in the form of business strategies for fundraisers.

Interestingly, the largest crowdfunding platforms in China are affiliated with the country's largest e-commerce platforms. These platforms, managed in the form of corporate entities, are primarily responsible for information management on the platform. This arrangement allows them to provide full link services, effectively transforming a good idea into a product and delivering it to users through the upstream and downstream supply chain of e-commerce and crowdfunding concept platforms \cite{funk2016institutions}.

However, this corporate management structure also means that the platform controls the crowdfunding ecosystem by managing information, leading to a lack of self-regulation mechanisms. Participants in these ecosystems cannot control their information, highlighting a key limitation of conventional crowdfunding platforms.

\subsubsection{Conventional Decentralized Crowdfunding Platform (ICO)}

\begin{figure}[htbp]
\centering
\includegraphics[width=1\linewidth]{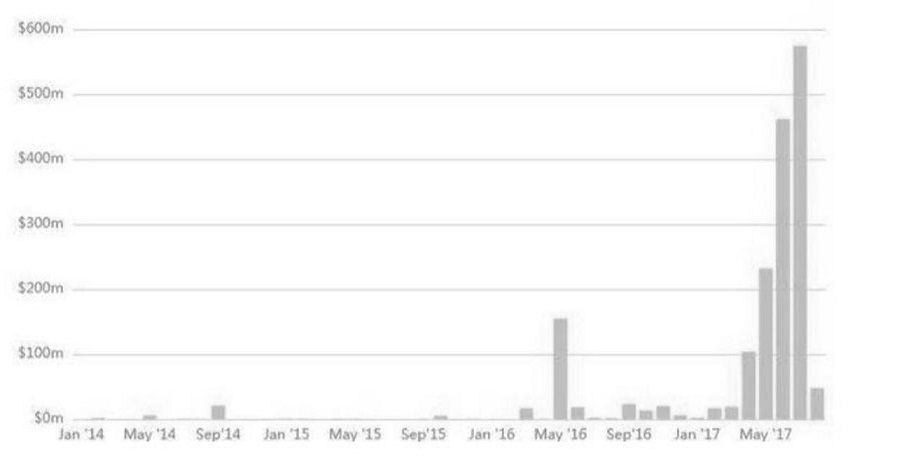}
\caption{Histogram of funds raised by ICO crowdfunding from 2013 to 2017}\label{fig3-1}
\end{figure}

Initial Coin Offering (ICO) crowdfunding, a form of crowdfunding economic activity that utilizes the trust mechanism of blockchain technology as a credibility carrier, emerged in 2013 and reached its peak in 2017 \cite{fridgen2018don}. In ICO crowdfunding, peer-to-peer transactions are conducted with the initial digital cryptocurrency generated in return (See Figure \ref{fig3-1}).

Table \ref{table3-2} outlines the main features of ICO crowdfunding platforms. According to Boreiko and Sahdev's research on the evolution of ICOs \cite{boreiko2018ico}, projects participating in crowdfunding through the ICO mechanism are gradually transitioning from individual projects to being part of a crowdfunding ecosystem consisting of clusters of crowdfunding projects.

\begin{table*}[htbp]
\centering
\includegraphics[width=1\linewidth]{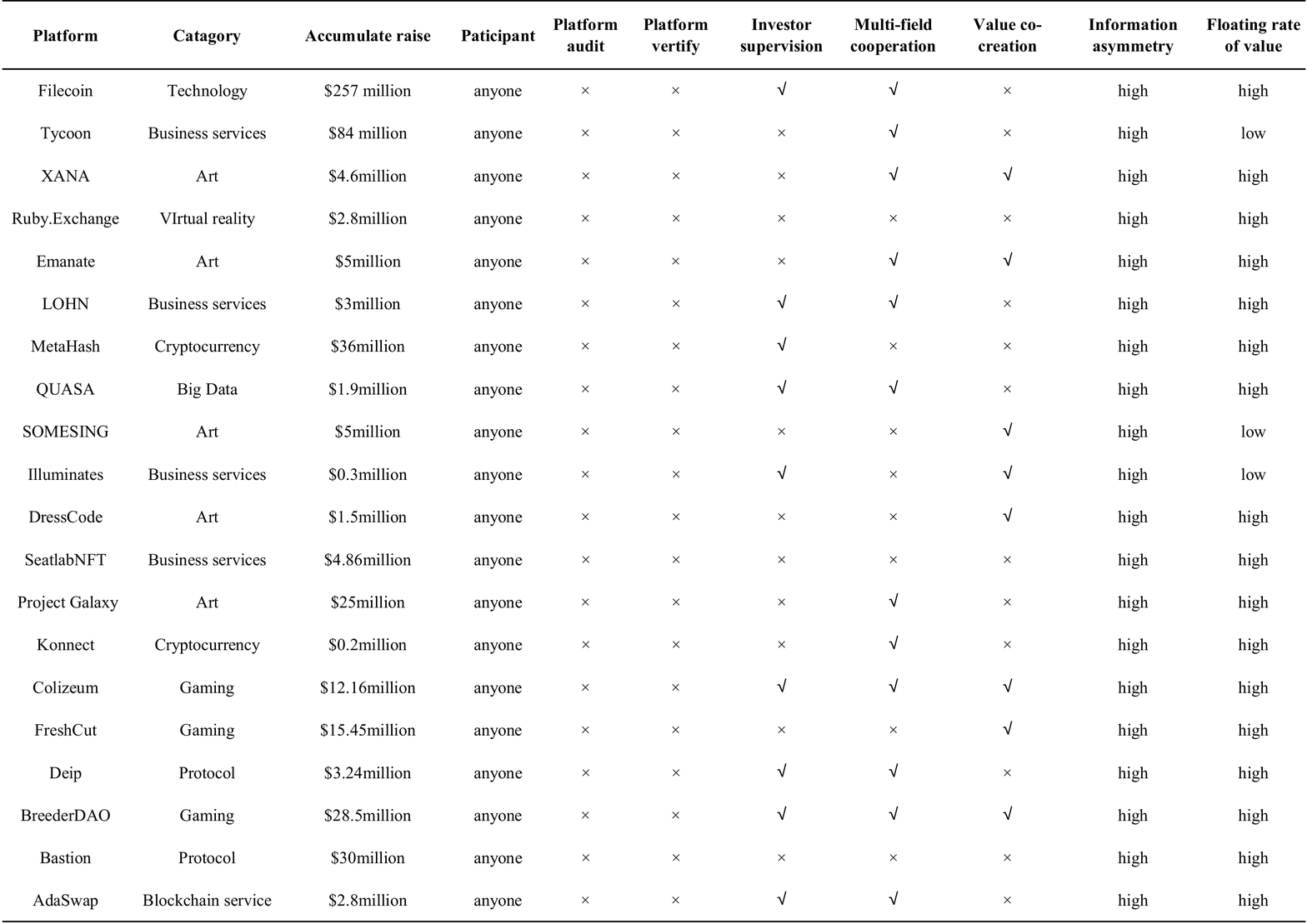}
\caption{ICO crowdfunding platforms}\label{table3-2}
\end{table*}

Initially, nearly two-thirds of ICO crowdfunding platforms engaged in multi-field cooperation, with a significant scale of participation in the form of project cluster ecosystems. Additionally, nearly a quarter of ICO crowdfunding platforms released their roadmaps. However, these crowdfunding projects lack content regulation and audit mechanisms, and a large percentage of them involve "air coins" \cite{arnold2019blockchain}.

The digital cryptocurrency used in ICO crowdfunding, built-in blockchain tokens, is primarily used for internal circulation and fee settlement on blockchain platforms. Its circulation scenario mainly involves the trading of virtual digital artworks, on-chain in-game products, and other virtual digital goods. This lack of association with the real economy can impact its stability and presents a high level of risk due to the unregulated and laissez-faire approach to economic activity in ICO crowdfunding mechanisms.

\subsubsection{Analysis of Stakeholders' Interest}

A comparison of Table \ref{table3-1} and Table \ref{table3-2} reveals significant differences between conventional crowdfunding and ICO crowdfunding. Firstly, most ICO crowdfunding platforms do not verify and supervise the projects on their platforms, while most conventional crowdfunding platforms have audit and oversight processes hosted by centralized platform company subjects. This results in a lower threshold for fundraisers on ICO crowdfunding platforms \cite{block2021entrepreneurial}. Secondly, ICO crowdfunding suffers from severe information asymmetry, while conventional crowdfunding platforms impose numerous restrictions on both participants (e.g., requiring fundraisers to be the legal representative of the project company and investors to show proof of funding source, etc.). Therefore, ICOs offer a higher degree of openness to both sides of the crowdfunding process.

\begin{figure*}[htbp]
\centering
\includegraphics[width=1\linewidth]{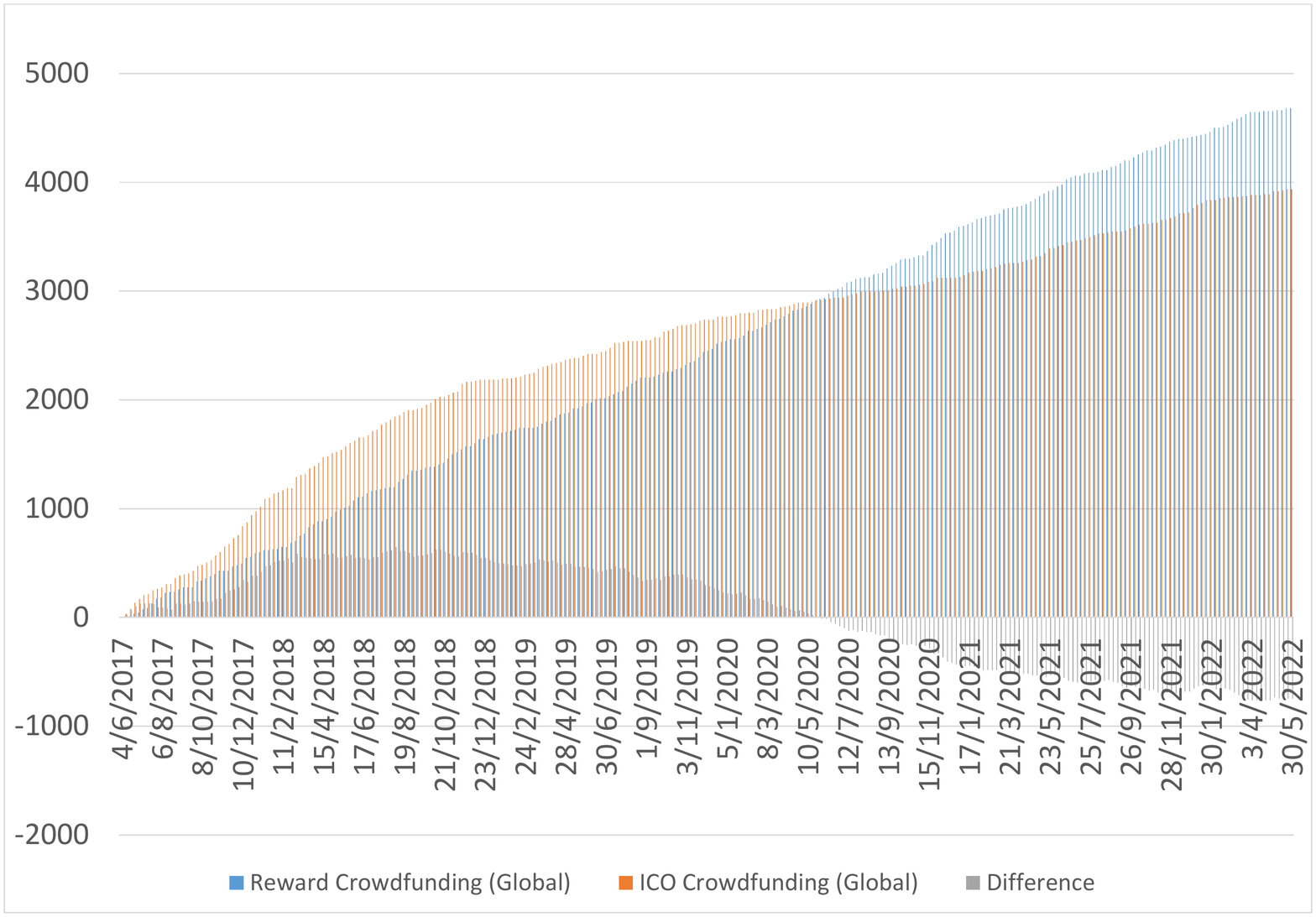}
\caption{Google search hotness of "ICO Crowdfunding" and "Reward Crowdfunding"}\label{fig3-2}
\end{figure*}

As shown in Figure \ref{fig3-2}, the ICO crowdfunding model has successfully attracted a larger group of people compared to conventional crowdfunding. The emergence of ICOs is accelerating the transformation of the economic structure towards a sharing economy, and has successfully shifted attention from conventional centralized crowdfunding to decentralized crowdfunding economic activity. The high openness and low threshold setting of ICOs, along with the curiosity of participants, led to an explosion in the popularity of the ICO mechanism in 2017 in terms of attention and accumulated funds. However, the rising risk of illegal fundraising due to its over-indulgence has led to increased insecurity, and the popularity of ICOs has begun to decline. In contrast, conventional crowdfunding, represented by reward crowdfunding, has remained stable and popular, relying on the steadiness of the real economy and the maturity of the ecosystem \cite{fisch2021motives}.

\subsection{Driving Role of Our Research Methodology}

The research methodology employed in this study plays a crucial role in the construction of our DCC model. By conducting a comprehensive analysis of both conventional centralized and decentralized crowdfunding platforms, we have been able to identify key strengths and weaknesses in existing models. This has informed the design of the DCC model, allowing us to incorporate features that enhance transparency, fairness, and shared governance, while mitigating common issues such as information asymmetry and high entry thresholds.

Our methodology also involved a detailed analysis of stakeholders' interests, which has been instrumental in shaping the tripartite division of communities within the DCC ecosystem. By understanding the distinct roles and responsibilities of different stakeholders, we have been able to design a system that encourages a well-rounded ecosystem, promotes checks and balances, and fosters a high degree of interaction and collaboration among the communities.

Furthermore, our legal and policy analysis has provided valuable insights into the regulatory landscape of crowdfunding, informing the design of the DCC model to ensure it aligns with legal requirements and best practices. This is particularly important given the evolving nature of regulations surrounding blockchain technology and crowdfunding activities.

\section{Modeling the DCC Ecosystem}

As the landscape of crowdfunding continues to evolve, the need for more innovative, inclusive, and equitable models is becoming increasingly apparent. The Decentralized Co-governance Crowdfunding (DCC) ecosystem stands as a promising response to these emerging needs. This section provides a comprehensive overview of the DCC ecosystem, exploring its unique token structure, community formation, project validation, and decentralized evaluation mechanisms. Inspired by the Steem and CGS but distinguished by its novel adaptations, the DCC ecosystem holds significant potential for driving a new era in crowdfunding – one that prioritizes transparency, fairness, and shared governance.

\subsection{DCC Ecosystem Communities}

The DCC ecosystem is structured around the principle of the tripartite division of communities, each representing distinct roles and responsibilities. This division ensures that the complex dynamics of the ecosystem, encompassing investment, content creation, and governance, are handled proficiently. It not only encourages a well-rounded ecosystem but also promotes checks and balances and fosters a high degree of interaction and collaboration among the communities.

\subsubsection{Capital Community: Coordinator of Resources}

The Capital Community comprises users holding the Capital Tokens, a relative stablecoin tethered to the value of gold with permissible fluctuations. Capital Tokens are envisioned as a robust and reliable asset within the ecosystem, maintaining their value despite the often tumultuous crypto markets. They can be obtained in exchange for real-world capital, assets, legal tender, crypto assets, or Labor Tokens.

The central role of the Capital Community is to drive investment and provide liquidity within the ecosystem. Their investments in Capital Tokens fund various crowdfunding projects, stimulating activity and fostering the growth of the ecosystem. Moreover, by providing necessary liquidity for these projects, the Capital Community ensures that resources can be swiftly allocated and reallocated according to the shifting needs of the ecosystem, thus promoting efficiency and flexibility. As investors, members of the Capital Community receive a return on their investments in the form of Capital Tokens, which further incentivizes active participation and contribution to the ecosystem's liquidity.

\subsubsection{Labor Community: Creators of Value}

\begin{figure*}[htbp]
	\centering
	\includegraphics[width=1\linewidth]{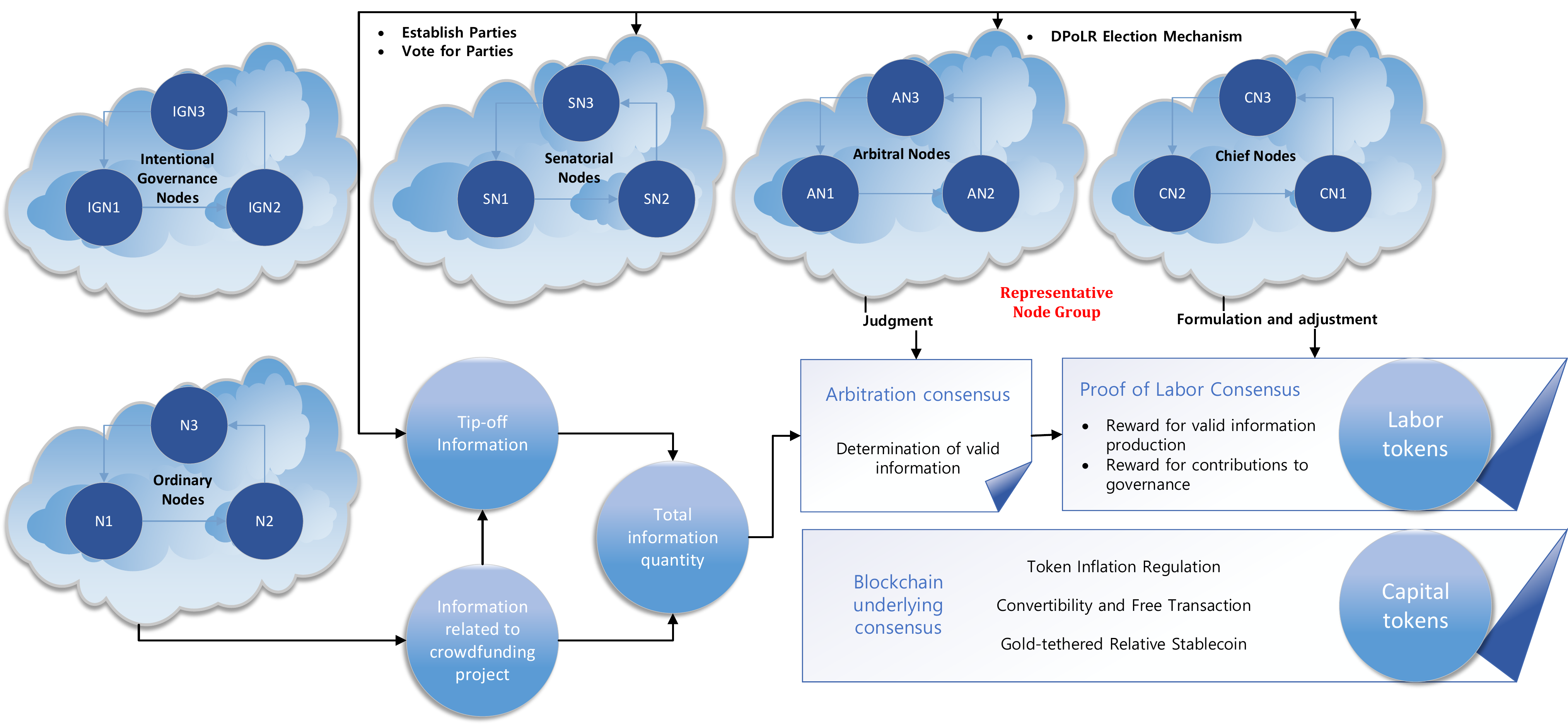}
	\caption{Production model of the tokens}\label{fig4-2}
\end{figure*} 

The Labor Community is a crucial part of the ecosystem, comprising users who earn Labor Tokens. These tokens are created whenever blocks containing valid content or governance information are produced, see Figure \ref{fig4-2}. A variety of activities can earn Labor Tokens in the ecosystem, such as creating valuable content, initiating crowdfunding projects, updating project processes, or releasing project outcomes and achievements. This system operates on the principles of the Proof of Labor consensus mechanism.

Labor Tokens are uniquely characterized by their unidirectional exchange mechanism, as per the primary smart contract of the ecosystem. Labor Tokens can be converted into Capital Tokens, demonstrating the monetization of labor within the ecosystem. Conversely, Capital Tokens cannot be directly converted back into Labor Tokens under normal circumstances, maintaining the integrity of labor contributions within the ecosystem.

However, there are two exceptions to this rule. Firstly, following successful crowdfunding, a portion of Capital Tokens obtained through the crowdfunding activity will be automatically converted back into Labor Tokens at the current exchange rate. This transformation occurs at each stage of project execution and serves as a reward for fundraisers to produce valid information in the forthcoming stage. Secondly, the gas fees paid by the Capital Community also get converted back to Labor Tokens at the current exchange rate and are transferred to the labor token reward pool.

The exchange rate between Labor Tokens and Capital Tokens isn't static but fluctuates in response to market conditions, ensuring that the value of labor aligns with the overall dynamics of the ecosystem. Additionally, Labor Tokens are not tradeable between users, further enhancing their role as an accurate reflection of individual user contributions.

By emphasizing value and content creation, the Labor Community is a key driver of the DCC ecosystem's vibrancy. The unique properties and restrictions tied to Labor Tokens uphold the principle of labor reward, ensuring that each token genuinely represents a user's contributions to the ecosystem.

\subsubsection{Governance Community: Custodians of the Ecosystem}

The Governance Community is composed of users holding Governance Tokens. Unlike other tokens, these cannot be generated directly. Users can obtain them only by converting their Labor Tokens, a process that is regulated and restricted by the primary smart contract of the ecosystem.

The conversion process from Labor Tokens to Governance Tokens is both delayed and phased. Once initiated, a specific portion of the Labor Tokens is converted into Governance Tokens at each phase until the conversion is complete. This gradual conversion process ensures that the allocation of governance power reflects consistent and ongoing contributions to the ecosystem. It also prevents sudden increase in governance power that might disrupt the ecosystem's stability and balance. On the contrary, the conversion from Governance Tokens back to Labor Tokens is instant, offering users flexibility when they wish to reduce their governance responsibilities or have urgent need on fund.

A notable restriction on Governance Tokens is that they are not allowed to be traded between users, like Labor Tokens. This rule further emphasizes the notion that governance power within the DCC ecosystem cannot be merely bought or sold, but must be earned through active labor contributions.

The quantity of Governance Tokens a user holds signifies their voting rights, demonstrating their willingness to participate in the governance of the ecosystem. This unique community plays a pivotal role in the DCC ecosystem's operations. From ecosystem governance and adjustment to project management and supervision, the Governance Community participates actively in decision-making processes, thereby helping steer the course of the ecosystem's development and assuring its integrity and fairness.

While each community has distinct roles and responsibilities, it's important to note that these communities are not mutually exclusive. Users may participate in multiple communities if they hold multiple types of tokens, fostering a robust, interdependent ecosystem. In this triad of communities, the system creates a balanced synergy where Capital holders provide resources, Labor contributors generate value, and Governance participants ensure the integrity and longevity of the ecosystem. This innovative DCC ecosystem embodies the principles of decentralized finance by promoting inclusive participation, fostering merit-based rewards, and ensuring system integrity through democratic governance. This novel approach is poised to offer new opportunities in the field of decentralized crowdfunding.

\subsection{Delegated Proof of Labor Representative (DPoLR) Election Consensus}

The Decentralized Co-governance Crowdfunding (DCC) ecosystem introduces a unique consensus mechanism—Delegated Proof of Labor Representative (DPoLR). This innovative approach allows creator nodes within the labor community to transform into intentional governance nodes, provided they convert their Labor Tokens into Governance Tokens. This transformation, firmly regulated by the ecosystem's primary smart contract, ensures a systematic and transparent transition from content creation to ecosystem governance.

\begin{figure*}[htbp]
\centering
\includegraphics[width=1\linewidth]{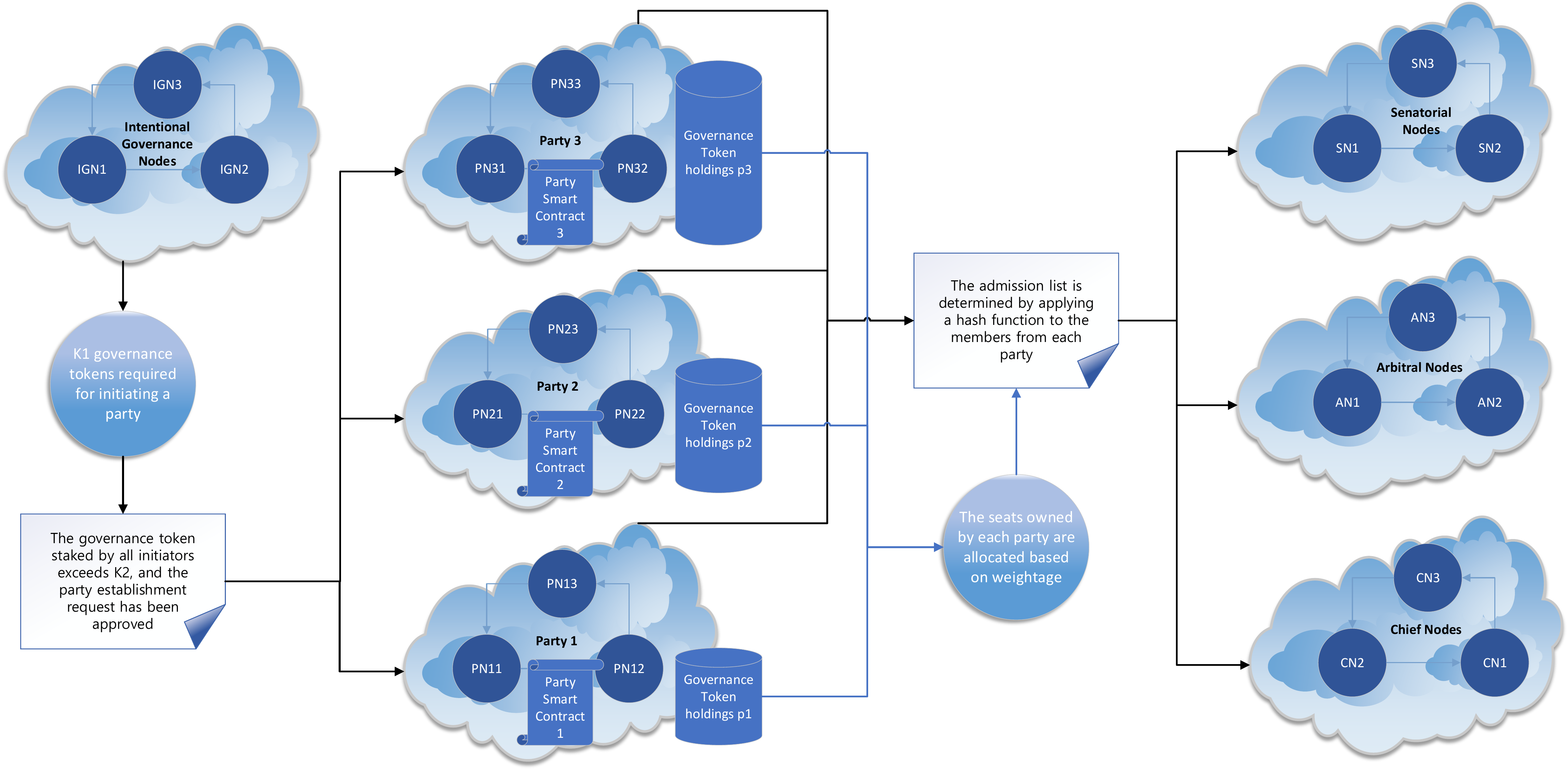}
\caption{Delegated Proof of Labor Representative Process}\label{fig4-10}
\end{figure*}

DPoLR then enables the formation of parties within the system. As shown in Figure \ref{fig4-10}, intentional governance nodes, upon amassing a particular quantity of Governance Tokens, have the privilege to establish parties and create party smart contracts, which function as their charters and guidelines. This way, they can articulate their party's ethos and operational framework. Other nodes select their affiliations based on alignment with these party smart contracts, ensuring ideological consistency within parties.

The influence a party wields within the ecosystem is a direct reflection of the total Governance Tokens its members hold. Importantly, a party's share of seats in the active governance community mirrors the proportion of Governance Tokens that party possesses. To ensure equitable representation, the specific party members eligible for these governance seats are determined through a hash function and rotate regularly. This process ensures all party members have an opportunity for election, though it's possible for lucky nodes to be chosen multiple times and thereby occupy multiple seats. In cases where nodes with significant Governance Tokens prefer not to affiliate with any party, they can form single-node parties and still actively participate in governance.

The Delegated Proof of Labor Representative consensus mechanism distinguishes itself from other models like Delegated Proof of Stake (DPoS) in several ways. Firstly, DPoLR promotes a distinct separation of capital and governance by allowing only Governance Tokens to dictate voting rights. By excluding Capital Tokens from voting, the system effectively curbs the influence of capital on elections, avoiding Matthew Effect and upholds relative fairness. Secondly, the adoption of the party system provides an expansive representation of diverse ideologies and interests within the ecosystem. This inclusivity encourages an active participation in governance and enriches the decision-making environment. Lastly, the use of hash functions in the selection of voting nodes ensures a process that is equitable, transparent, and resistant to manipulation. This guarantees every party member a chance at being elected, reinforcing a sense of fairness and inclusivity within the system.

In sum, the DPoLR consensus mechanism, which uniquely merges technological advancements with principles of political science, stands as a distinctive feature of the DCC ecosystem. It aims to balance power, foster fair representation, and create a decentralized, co-governed environment that is truly of the user, by the user, and for the user.

\subsection{Tripartite Governance Architecture to Achieve Benefit-sharing}

\begin{figure*}[htbp]
\centering
\includegraphics[width=0.8\linewidth]{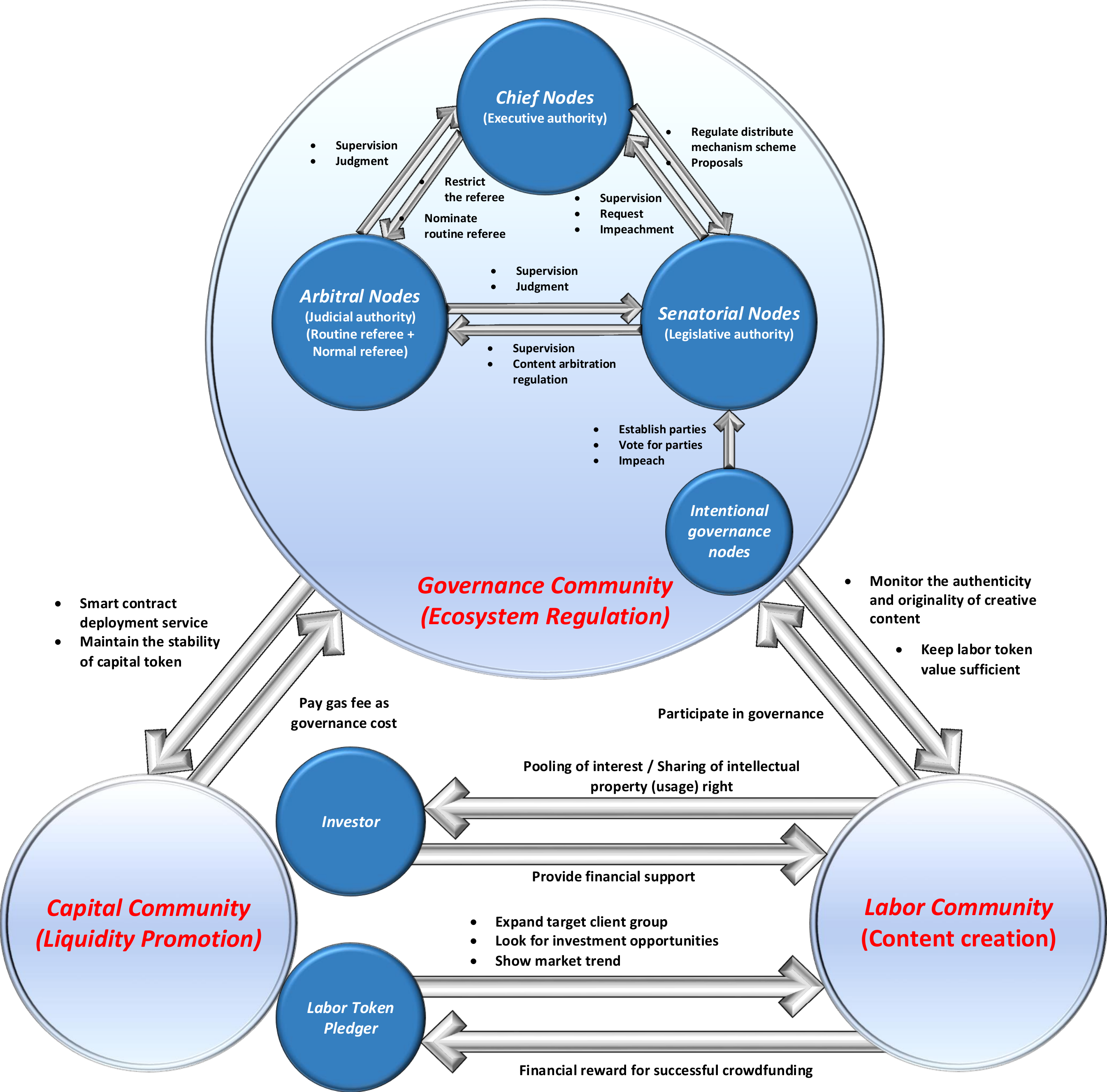}
\caption{Crowdfunding community structure}\label{fig4-9}
\end{figure*}

The Decentralized Co-governance Crowdfunding (DCC) ecosystem establishes an innovative tripartite governance architecture, bringing together the governance community, labor community, and capital community to realize the proposed crowdfunding system's consensus mechanism, see Figure \ref{fig4-9}. This mechanism ensures the equilibrium of rights and responsibilities across these communities while maintaining the robustness and self-sustainability of the ecosystem.

The labor community serves as the birthplace of information and content and initiates funding for new ideas. The capital community, divided into investors and pledgers, fuels these ideas with their resources. Investors contribute with legal tender or capital tokens to provide financial backing for creators, reaping shared interest and intellectual property rights as rewards. Pledgers, on the other hand, endorse the projects with labor tokens. By helping creators broaden their target client group, identify investment opportunities, and demonstrate market trends, pledgers also partake in the potential success of the projects, earning capital tokens as marketing rewards.

The governance community, formed in alignment with the political theory of the separation of powers, plays a crucial role in this ecosystem's architecture. This community follows the principle of separated governance, providing a model that ensures checks and balances while fostering autonomy and cooperation. The governance community's structure is composed of active governance nodes, which are elected from a diverse array of parties. These parties, in turn, are formed by intentional governance nodes who have opted to actively partake in decision-making and are elected via DPoLR mechanism.

The active governance nodes fall into three categories: chief nodes, senatorial nodes, and arbitral nodes, each carrying distinctive responsibilities and authorities. Chief nodes primarily adjust the incentive pool, orchestrating the allocation of labor tokens between the labor community and the governance community. In addition to this, they possess the authority to modify parameters, like Labor to Governance token exchange parameter, within the ecosystem, thereby exercising administrative power. Senatorial nodes hold the primary responsibility for formulating and amending the ecosystem's underlying smart contracts, thus possessing legislative authority. They shape the rules that govern the system, enabling flexibility and adaptability in the face of evolving needs and conditions. The arbitral nodes, on the other hand, are responsible for adjudicating creative content or reported information. They package blocks, witness block generation, and exercise judicial authority, ensuring fairness and accountability in the ecosystem.

In the interplay of governance, each node category has distinct rights and responsibilities. Chief nodes can propose smart contract modifications to senatorial nodes and have the right to nominate the next round of arbitral nodes. During subsequent node elections, the smart contract triggers a hash function that adjusts to increase the chances of nominated candidate arbitral nodes being elected. Meanwhile, senatorial nodes can request parameter modifications from the chief node and supervise or even impeach the chief node if necessary. They also oversee arbitral nodes and develop arbitration rules for them. Arbitral nodes maintain the right to supervise and judge both chief nodes and senatorial nodes, ensuring a balanced and accountable system of governance.

The governance community also provides vital services to the capital and labor communities. It provides smart contract deployment services and maintains capital token stability for the capital community, receiving gas fees in return, and monitors the authenticity and originality of creative content and ensure labor token value sufficiency for the labor community, Thereby ensuring the capital community's trust and preserving the credibility and sustainability of the labor community's rewards system. Through this intricate system of roles, responsibilities, and checks and balances, the governance community ensures the smooth operation of the entire DCC ecosystem.

In this structurally stable architecture, the incentive consensus guided by governance sharing undergoes self-adjustment. To actualize community value benefit-sharing, the ecosystem's overall benefits must align with each community's interests. The foundational incentive distribution adjustment mechanism in the blockchain crowdfunding system aims to balance the total value rewards for each community, thereby establishing a harmonious, decentralized, and co-governed environment.

\subsection{Transformation of Crowdfunding Ecosystem Governance Mechanism}

The proposed co-governance mechanism for the crowdfunding ecosystem envisages a shift from centralized to decentralized governance. However, it's important to acknowledge the necessity of initiating this system from a centralized platform. It's not feasible to jump directly into a fully Decentralized Co-governance Crowdfunding (DCC) system due to several practical considerations, the foremost being the need for a substantial number of nodes to facilitate effective decentralized co-governance. In the early stages, the ecosystem is unlikely to possess enough nodes to maintain a truly decentralized platform. Hence, a centralized structure is a pragmatic and temporary solution.

Conventional centralized crowdfunding platforms rely on organizational structures that involve manual audits by full-time employees, often supplemented with evaluations from centralized organizations like venture capital firms \cite{michels2012unverifiable}. While this method serves its purpose, it often neglects the influence of group consensus information provided by social media, which can be a significant factor in determining the popularity of products \cite{saxton2014social}. This inherent limitation underscores the need for a transition towards a more decentralized system that incorporates broad community feedback.

As the number of nodes participating in the ecosystem increases, the platform can begin to shift towards a more decentralized form of governance. Nodes are recruited to participate in supervision through a carefully structured incentive mechanism, and chief nodes are selected from among the intentional governance nodes to participate in adjudication using the DPoLR mechanism. This incentive system serves to boost the enthusiasm of nodes to engage in and supervise project information.

During this transitional period, the platform gradually reduces the proportion of centralized supervision nodes that it hosts, handing over more control to the decentralized community. A significant aspect of this transition involves the establishment of a supervision incentive mechanism. This mechanism aims to stimulate nodes to actively participate in and oversee project information, driving the decentralized transformation of the platform's supervision system.

This transformation from a centralized to a decentralized system doesn't happen overnight. It's a phased process that relies on growing participation and the gradual redistribution of control. The idea is not to discard the useful aspects of centralized structures but rather to merge them with the advantages of decentralized systems. Over time, as the ecosystem evolves and matures, the platform becomes a fully decentralized co-governance mechanism that incorporates diverse inputs and maintains robust checks and balances. In this manner, the crowdfunding ecosystem can optimize project supervision and enhance community participation, fostering a more democratic and efficient operation.

\section{Risk Mitigation Strategies in DCC Activities}

In any financial ecosystem, particularly those involving investments and crowdfunding, risk management is a critical aspect that ensures sustainability and investor confidence. In a Decentralized Co-governance Crowdfunding (DCC) ecosystem, the decentralized nature and shared governance present unique challenges and opportunities for risk control. This section delves into these aspects, discussing the potential risks inherent to the DCC system and presenting strategies to mitigate these risks. The objective is to ensure that the DCC ecosystem remains resilient and reliable, preserving its innovative approach to crowdfunding while safeguarding the interests of all participants.

\subsection{Blockchain-Based Identity Authentication and Regulatory Compliance}

\begin{figure*}[htbp]
	\centering
	\includegraphics[width=1\linewidth]{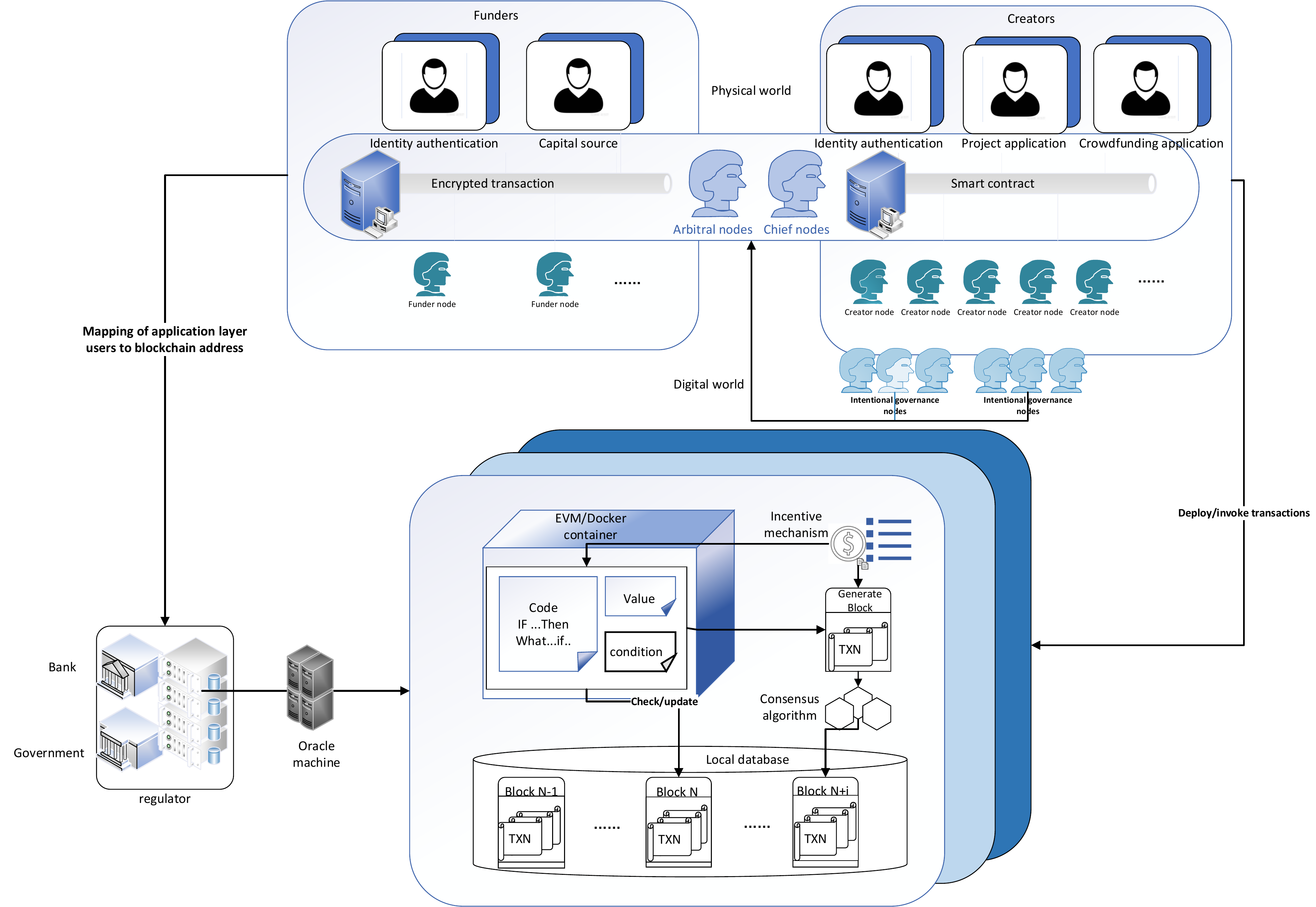}
	\caption{Intelligent identity authentication system}\label{fig4-1}
\end{figure*}

Crowdfunding platforms operating within the real economy are obligated to adhere to the legal provisions set forth by regional regulatory authorities \cite{authority2014fca}. As illustrated in Figure \ref{fig4-1}, a blockchain-based crowdfunding system necessitates the establishment of a robust user authentication center. This center is responsible for managing the generation of public and private keys, which are integral components of the blockchain infrastructure.
A fundamental prerequisite for crowdfunding is the supervision and management of identity information by regional regulatory agencies. This ensures that all transactions and interactions within the platform are conducted in a transparent and accountable manner, thereby fostering trust among users and stakeholders.

The core technology of blockchain extensions, the smart contract, plays a pivotal role in this context. Smart contracts serve as automated digital intermediaries that facilitate, verify, and enforce the negotiation or performance of a contract. In the context of a crowdfunding platform, they play a crucial role in managing the diverse and complex information associated with creative projects. For funders, who come with a range of qualifications and credit histories, smart contracts provide a streamlined and efficient way to handle their contributions. They automate the process of matching funders' qualifications and credit status with suitable creative projects, thereby ensuring that every investment is well-aligned with the funder's capabilities and risk tolerance. For creators, smart contracts offer a customizable framework that caters to their unique preferences and requirements. Whether it's setting specific funding goals, defining the terms of fund usage, or establishing criteria for funder participation, smart contracts can be tailored to meet the distinct needs of each creator. This flexibility not only enhances creators' willingness to disclose information but also mitigates project risks and establishes a high rank of credit. Furthermore, smart contracts can be instrumental in resolving disputes, thereby contributing to the overall integrity and reliability of the platform.

\subsection{On-Chain Protection Mechanisms for Smart Contracts}

\begin{figure*}[htbp]
	\centering
	\includegraphics[width=1\linewidth]{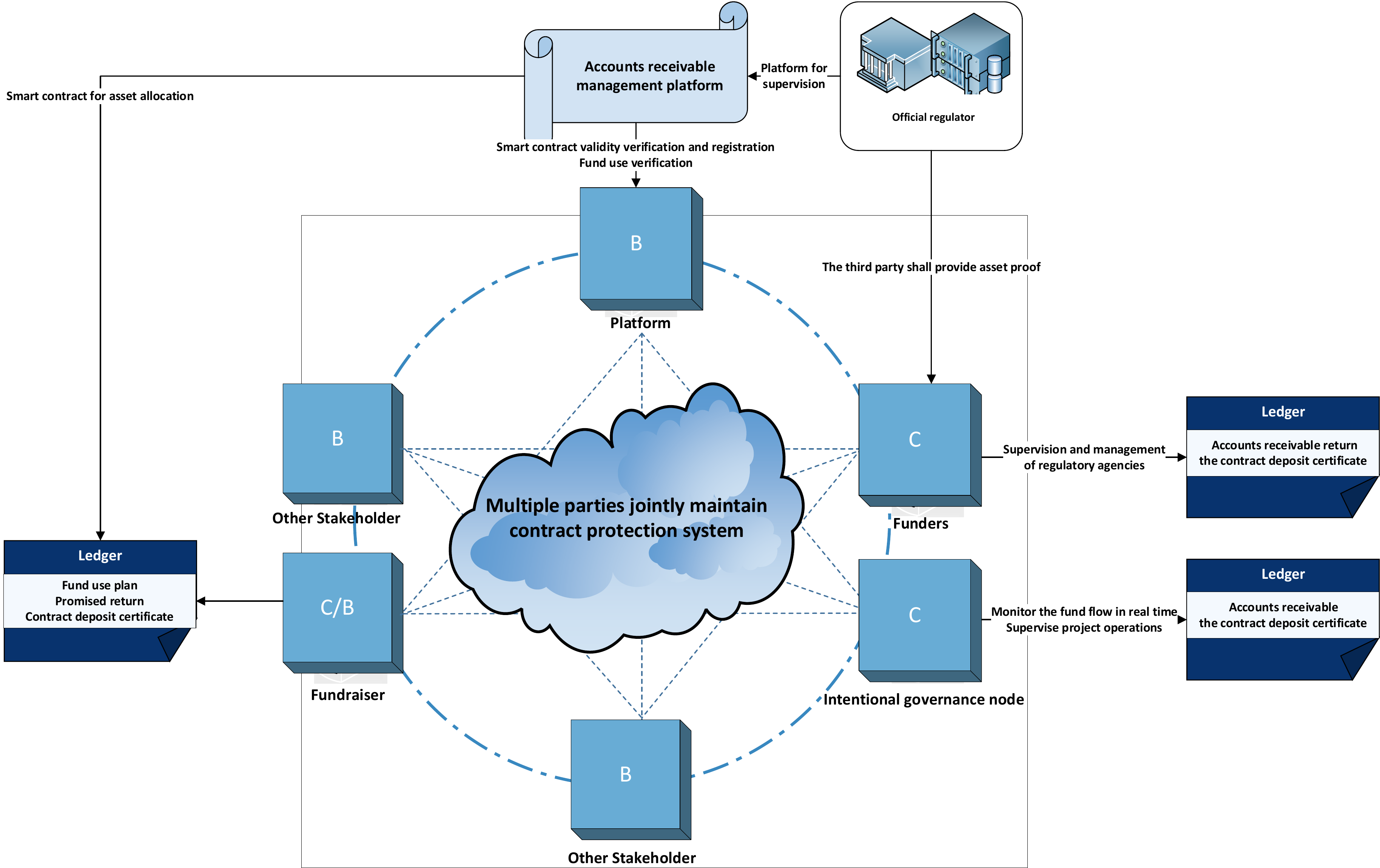}
	\caption{On-chain contract protection model}\label{fig4-4}
\end{figure*}

Given the research on security vulnerabilities in Ethereum blockchain smart contracts \cite{kushwaha2022systematic}, it's crucial to ensure the secure execution of crowdfunding processes. To successfully raise funds within the target scope, it's necessary to issue contract documents such as the "Fund Use Plan" and "Reward Contract" on the blockchain. This approach not only bolsters the security of these contract documents but also provides more reliable legal evidence, as depicted in Figure \ref{fig4-4}.

These on-chain documents are made accessible to all participants for review and evaluation, fostering transparency and trust within the crowdfunding platform. The platform only grants the project permission to access the raised funds once the contracts have been accepted by a certain portion of funders. This process ensures that the funds are used as intended and in accordance with the agreed-upon terms, thereby fostering trust among all participants and stakeholders. If the contracts are not accepted, they need to be reformulated to better align with the funders' expectations and the project's requirements.

In addition to the primary crowdfunding process, fundraisers can also take advantage of additional financial services offered by stakeholders. Any contracts signed with these stakeholders are also recorded on the blockchain. This on-chain record-keeping further enhances the security and transparency of the platform, providing a robust and reliable environment for both creators and funders. The on-chain protection of smart contracts ensures that all transactions and agreements are transparent, verifiable, and immutable, reinforcing the platform's trustworthiness.

\subsection{Decentralized Autonomous Council-Centered Architecture}

Crowdfunding platforms have revolutionized the way creators bring their ideas to life, leveraging the power of the internet to connect with potential funders. These platforms enable companies to garner attention, build trust, and secure necessary funding for their projects through direct engagement with their supporters. However, significant challenges such as fraud, limited user control, and information asymmetry pose substantial hurdles for both conventional centralized crowdfunding platforms and decentralized ones (ICOs), primarily due to the lack of a decentralized trust mechanism \cite{belleflamme2015economics}.

To address these challenges, we propose a composite decentralized co-governance crowdfunding architecture. This architecture, inspired by the DAICO system, represents an advanced version that addresses its limitations and enhances its strengths. It uniquely combines the project participation and management mechanisms of conventional centralized crowdfunding platforms with the inherent spirit of decentralization found in ICOs. This integration forms a robust framework where all users who contribute funds or participate in the credit scoring system can join the decentralized autonomous council of the project, playing a crucial role in overseeing the project's operations and risk management. The extent of each council member's supervisory authority over the project is determined by a combination of their investment amount and their willingness to participate in governance.

Council is composed of funders of the project and supervisors from Arbitral Nodes. In the event of project anomalies or risks, council members can promptly raise alerts and propose necessary course corrections through a feedback system enabled by smart contracts. This feedback from the mass supervision system not only ensures the project's accountability but also directly influences the project's quality improvement. If the project does not respond appropriately, the distribution of funds can be suspended, thereby mitigating potential risks.

\begin{figure*}[htbp]
	\centering
	\includegraphics[width=1\linewidth]{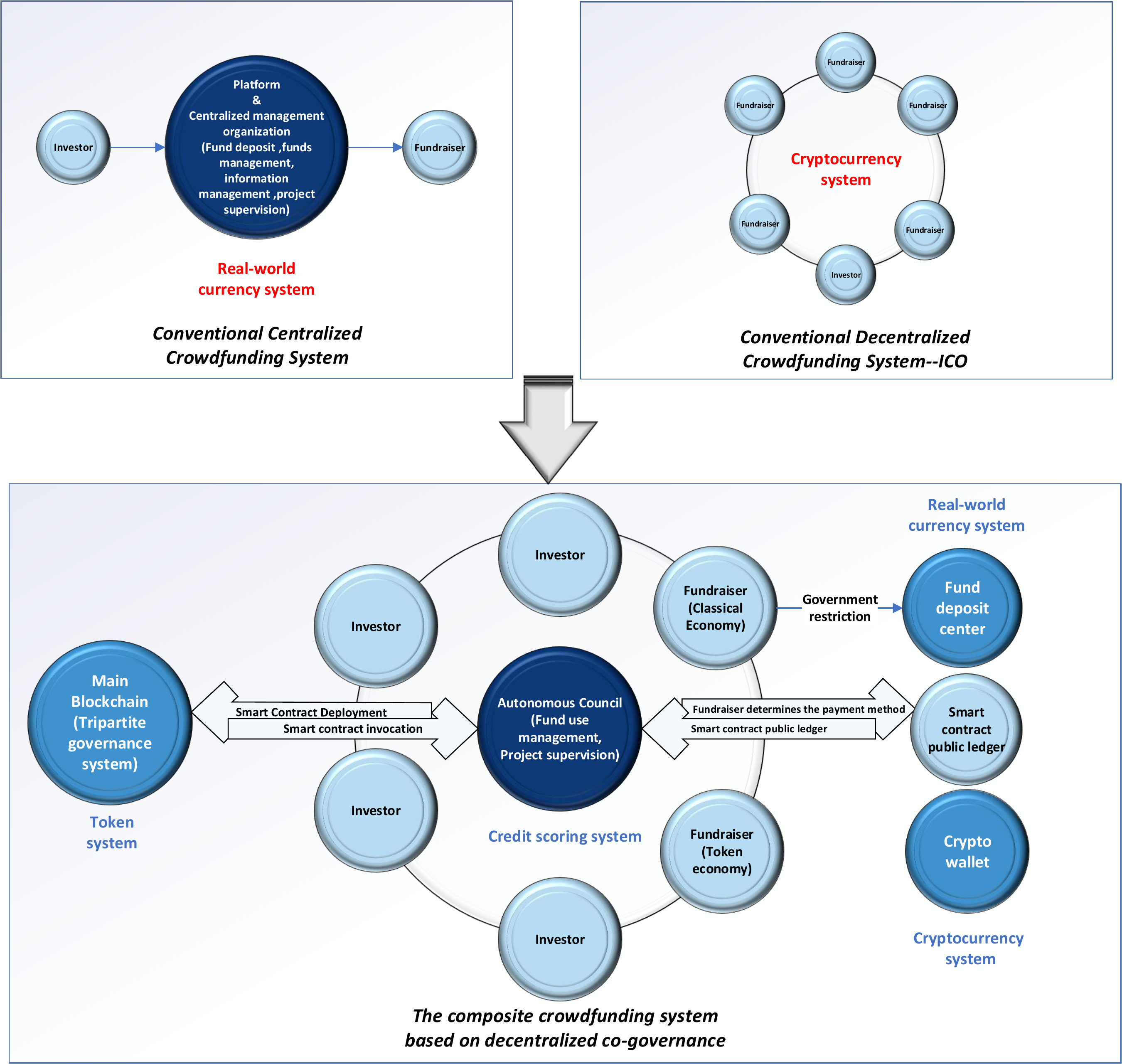}
	\caption{Comparison between DCC model and existing crowdfunding models}\label{fig4-6}
\end{figure*}

By implementing this composite decentralized co-governance crowdfunding architecture, we aim to significantly reduce the risks associated with incomplete crowdfunding projects prevalent in conventional platforms. Furthermore, our approach enhances project execution, management, and ensures the rights of funders. As depicted in Figure \ref{fig4-6}, this composite decentralized co-governance crowdfunding architecture offers a promising solution to the challenges and risks faced by conventional crowdfunding platforms.

\subsection{Distributed Token Evaluation System for Concept Validation}

\begin{figure*}[htbp]
	\centering
	\includegraphics[width=1\linewidth]{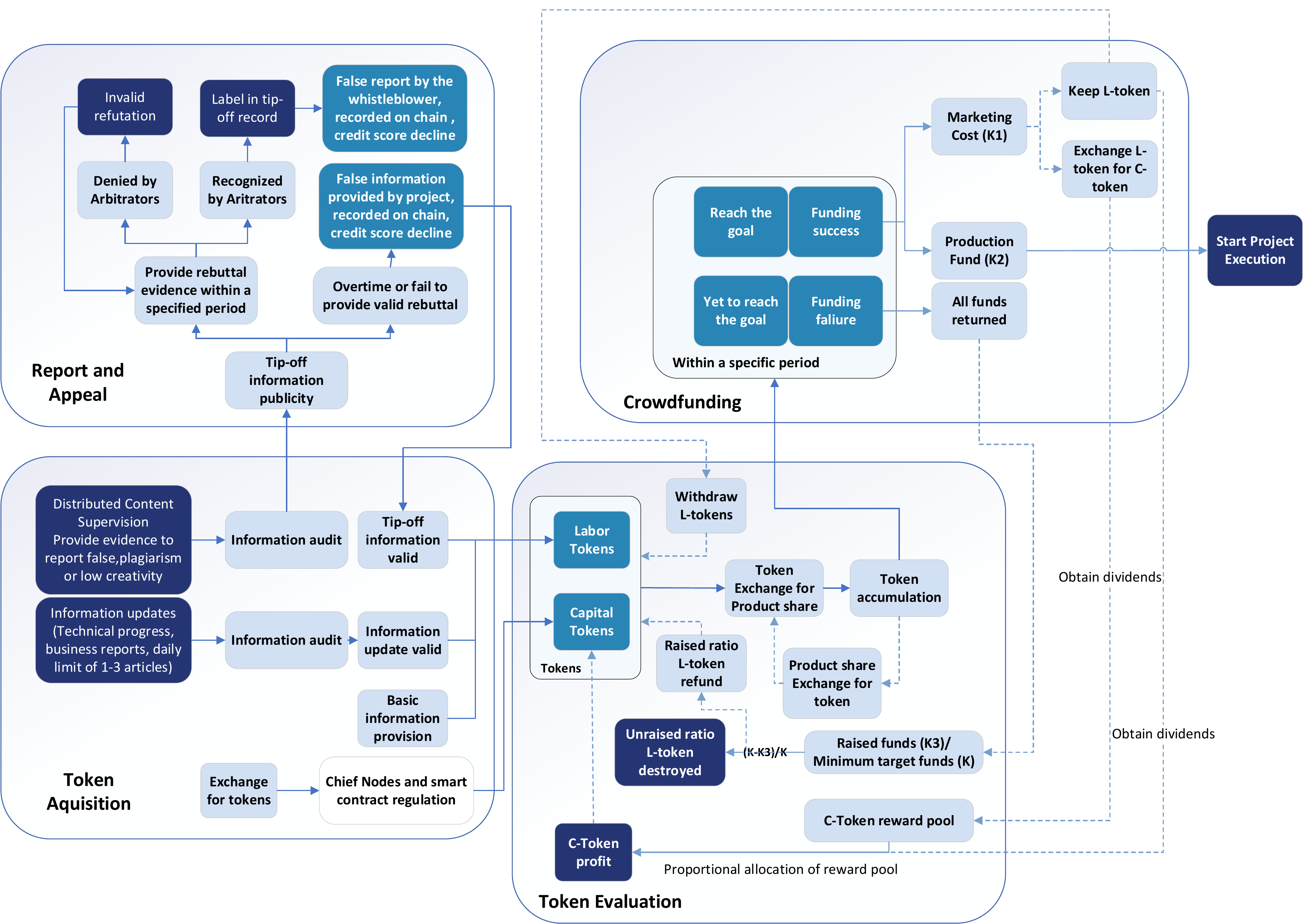}
	\caption{Token generation and circulation modes}\label{fig4-3}
\end{figure*}

In the conventional crowdfunding paradigm, as pointed out by Jensen\cite{jensen2002value}, a centralized platform functions similarly to a corporation, increasing the company's value. However, this centralized evaluation process can suffer from moral hazard issues when the platform and fundraisers become tightly interlinked stakeholders. The centralized nature of the system might leave funders with limited information, thereby increasing the probability of infringements on their rights and interests. Conversely, a decentralized evaluation system facilitated by blockchain technology can circumvent these shortcomings, providing a fair and transparent review framework. As shown in the right side of Figure \ref{fig4-3}, to fulfill the requirements of crowdfunding for creative projects, the ability to transparently demonstrate the merits of their concept to funders is paramount. In this context, the distributed token evaluation system becomes crucial, enabling funders to acquire unbiased reviews of the financing projects.

In this decentralized system, token rewards serve as incentives in reward crowdfunding when used as a circulation tool in the market \cite{kuppuswamy2018crowdfunding}. Unlike conventional reward crowdfunding, which merely arouses the spirit or belief in creators, blockchain tokens can help investors decide which projects to support and also enhance their visibility as part of project marketing teams. These tokens also serve as evaluation tools that creators offer through marketing to consumer groups participating in project crowdfunding.

Token evaluation groups, earning Labor Tokens through previous creative activities, pledge these tokens on projects they deem worthy. These pledged tokens act as a benchmark for the project's acceptance by potential investors or customers, symbolizing the target customer group's approval of the project concept and incentivizing fundraisers. By pledging on the project, token evaluators express their endorsement for the venture. According to Mitchell's theory of stakeholder benefit-driven development\cite{mitchell1997toward}, token evaluators, while seeking to maximize self-interest, will use their tokens with consideration for the potential returns on their token investment.

Tokens can also act as a risk-hedging means for the crowdfunding ecosystem, addressing the issue of information trust \cite{gatautis2014crowdfunding}. This role complements the participation of funders in crowdfunding and artists' fundraising. These risk management tools can successfully close the psychological gap between those involved in economic activity. When combined with the AON or KIA model of crowdfunding, tokens can also be utilized to obtain income from the marketing work required for the success of crowdfunding projects \cite{cumming2020crowdfunding}.

Creative projects within this ecosystem seek funding in the form of Capital Tokens or legal tender. When setting up a project, the fundraiser stipulates a marketing reward proportion in advance, which determines how the capital raised will be allocated. Should the project successfully reach its fundraising goal by raising the requisite amount of Capital Tokens, the pre-determined proportion of Capital Tokens is distributed to token evaluators as a marketing reward. The remaining Capital Tokens serve as the production fund, which undergirds the operational aspect of the project. Additionally, all the Labor Tokens pledged are refunded to their respective contributors.

Each pledger's share of the marketing rewards mirrors the proportion of their pledged Labor Tokens to the total Labor Tokens pledged to the project. However, if a project fails to amass enough Capital Tokens within the specified time, the raised Capital Tokens are returned to the contributors, and a proportion of the Labor Tokens are refunded, with the rest being destroyed. The refund rate for Capital Tokens mirrors the funded rate. For instance, if a project has raised 40 tokens out of a target of 100 tokens when the deadline hits,the funded rate stands at 40 percent. Consequently, 40 percent of the Labor Tokens will be refunded to token evaluators, and the remaining 60 percent will be transferred to an inaccessible wallet address and destroyed.

The token evaluation group can function as marketers promoting the project to raise production capital, or as investment advisors collaborating with funders. The growth of tokens reflects the marketing intentions of the marketing groups and the market value evaluation of a product. The token income symbolizes the return on the marketing groups' contribution to successful crowdfunding and can also be viewed as the dividend earned by aiding crowdfunding groups in making profitable investments.

This model for project concept validation and evaluation provides a harmonious compromise between the All-or-Nothing (AON) and Keep-it-All (KIA) models. While not strictly adopting the AON model, it allows for a partial refund of Labor Tokens if a project fails, thus mitigating financial risks to participants. Similarly, it doesn't entirely conform to the KIA model as a proportion of Labor Tokens are destroyed if the project doesn't achieve its funding goal. This ensures a balanced distribution of risks and rewards, upholding the principles of a decentralized co-governance system. Simultaneously, it serves to incentivize token evaluators by providing potential dividends for successful crowdfunding campaigns, effectively rewarding their contribution and investment in project evaluation and promotion.

The innovation of the distributed token evaluation system lies in creating a token evaluation group, bridging the gap between fundraisers and funders, enabling a virtuous cycle between creative projects and funders. Projects are incentivized to improve their creative value to garner more support, while funders can regard tokens as a reference for market prospects. This system not only helps investors decide which projects to support but also makes them more visible as part of project marketing teams, thereby closing the psychological gap between those involved in economic activity.

\subsection{Distributed and Token-Incentivized Supervision and Credit Scoring System}

The Decentralized Co-governance Crowdfunding (DCC) ecosystem introduces a token-incentivized distributed supervision and credit scoring system, which combines a distributed information audit mechanism and a satisfaction feedback system. This strategy promotes information accuracy, reliable participation from users, and effective project risk control through a blend of token incentives, community arbitration, and collective reputation\cite{agrawal2014some}.

A fundamental component of this system involves the preliminary audit by designated arbitral nodes. These nodes apply evaluation criteria within predefined periods known as "evaluation intervals" to review submitted reports. Rather than basing the report's validity on a single evaluation, the system utilizes a cumulative approach, tallying multiple evaluations over a set period. This decentralized method aims to distribute responsibilities, supervision, and judicial powers within the ecosystem. When the blockchain system verifies reporting information as valid, it accordingly disburses incentive Labor Tokens following a pre-established mechanism.

'Creator nodes' within the labor community play a significant role by contributing information through valuable content, initiating crowdfunding projects, updating project processes, or releasing project outcomes and achievements. Upon passing the preliminary audit, the blockchain records this information and awards Labor Tokens to the creator nodes. However, if future reports successfully flag the submitted creative information as plagiarized, fake, fraudulent, or of low-creativity, the system will forcibly destroy the rewarded Labor Tokens and lower the creator's credit score, upholding the ecosystem's information integrity and credibility.

The mechanism further extends to nodes with qualifying credit scores, allowing them to propose a tip-off for issues such as plagiarism, fakeness, fraud, or low-creativity. See the left side of Figure \ref{fig4-3} Once the preliminary audit by arbitral nodes is cleared, the system promulgates this tip-off information. The accused project is then mandated to provide rebuttal evidence within a specific timeframe. If the arbitration community accepts the evidence, the system stamps the tip-off information with "refute successfully" and records any suspected false report by the whistleblower on the blockchain, deducting the whistleblower's credit score. Conversely, if a project fails to provide valid refutation by the deadline, the system records the project's suspected provision of false information on the blockchain, deducts the project's credit score, and awards the whistleblower with Labor Tokens.

As highlighted by Winfree et al.'s study on product quality and reputation \cite{winfree2005collective}, a reputation mechanism that utilizes collective reputation and public evaluation can effectively regulate the interests and morality of fundraisers. Therefore, establishing a satisfaction feedback system for funders to evaluate the effectiveness of fundraisers in using and managing the funds they have raised is essential. This system serves as a vital tool for building a fundraiser's reputation during the fundraising process.

\begin{figure*}[htbp]
	\centering
	\includegraphics[width=1\linewidth]{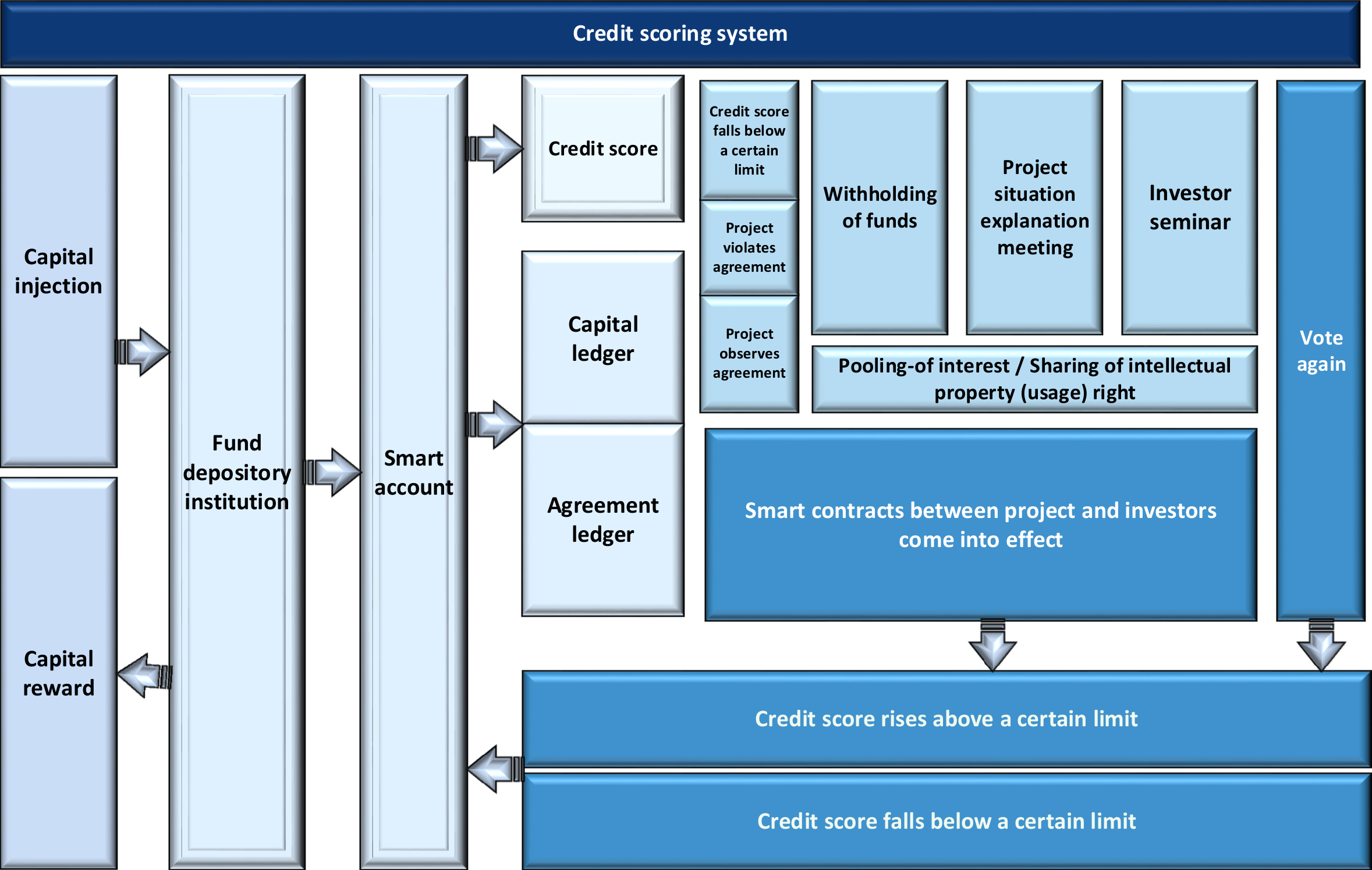}
	\caption{Overview of the credit scoring system}\label{fig4-5}
\end{figure*}

A project's credit score can decrease due to ambiguous usage of project funds, failure to transparently disclose the operating status, and delays in project delivery. As depicted in Figure \ref{fig4-5}, if funders discover that a project is not adhering to the agreed-upon terms, they can express their dissatisfaction by lowering the project's credit score. When the score reaches a warning line, the administrative nodes in the autonomous council will intervene in the investigation and further lower the credit rating. If the credit rating falls too low, the project may be suspended or denied access to project funds.

The process of altering credit scores, along with the documentation and fund flow, are all recorded on an immutable blockchain. The introduction of the satisfaction feedback and credit scoring system enables dynamic project oversight. The satisfaction of funders and the use of funds will continually impact the project's credit score. In addition to tracking the flow of funds, the credit score reputation system developed by the blockchain P2P network \cite{nakamoto2008bitcoin} enables point-to-point effective recording and comprehensive oversight between funders andfundraisers. This system ensures transparency, accountability, and trust in the crowdfunding process, thereby significantly enhancing project risk control.

As illustrated in Figure \ref{fig4-3}, the decentralized structure of the crowdfunding ecosystem equips each user node with the right to publish supervision (tip-off) and report proposals, which are then audited by arbitral nodes. By employing a distributed group of arbitral nodes, the system harnesses multiple evaluation criteria to determine the report's validity, effectively decentralizing responsibility, supervision power, and judicial power sharing. This expansion of the supervision group to encompass the entire user group amplifies the original supervision authority. Consequently, every independent node becomes a force for maintaining the stable operation of the crowdfunding ecosystem, enabling group auditing under a decentralized structure and showcasing the potency of decentralization in ensuring transparency, accuracy, and reliability in information audit and governance.

\section{Economic Analysis of the Sustainability of DCC Ecosystem}

In this section, we delve into the economic principles that underpin the sustainable development of the Decentralized Co-governance Crowdfunding (DCC) ecosystem. By leveraging economic theories and models, we aim to provide a comprehensive understanding of how this novel ecosystem can maintain its growth and stability over time. This analysis will shed light on the potential of DCC to revolutionize the way projects are funded and managed, offering a sustainable and efficient alternative to conventional methods.

\subsection{Game Behavior under Decentralized Co-governance}

Game theory, specifically the concepts of Nash Equilibrium and Pareto Optimality, offers a valuable perspective for analyzing the behavior of the stakeholders within a Decentralized Co-governance Crowdfunding (DCC) ecosystem.

\subsubsection{Nash Equilibrium in Ecosystem Communities}

The DCC ecosystem is a dynamic and interactive environment where the Capital Community, Labor Community, and Governance Community work together to create value, maintain stability, and ensure fair distribution of benefits. This interplay can be understood through the lens of game theory, specifically the concept of Nash Equilibrium.

In the DCC ecosystem, these roles are represented by the Labor Community (fundraisers), the Capital Community (investors), and the Governance Community (platform managers). Capital Community (C) invests in Capital Tokens to fund various projects, Labor Community (L) contributes by creating valuable content and earning Labor Tokens, and Governance Community (G) oversees the ecosystem's operations and maintains its integrity. Each community has two strategies: to invest/contribute/govern (I) or not to invest/not contribute/not govern (N). The Nash Equilibrium in this context would be a state where all three communities choose to invest/contribute/govern (I), as this strategy yields the highest benefits for all. This can be demonstrated through the following scenarios:

Reducing Risks for Fundraisers (Labor Community): In the DCC ecosystem, the Labor Community creates valuable content and initiates crowdfunding projects. They are incentivized by Labor Tokens, which they earn for their contributions. This system reduces the risk of excessive disclosure of project information, as the value of Labor Tokens is tied to the actual contributions of the Labor Community, not the amount of information they disclose.

Mitigating Risks for Investors (Capital Community): The Capital Community invests in Capital Tokens to fund projects. These tokens maintain their value despite market fluctuations, reducing investment risk. Moreover, the transparency and decentralization of the DCC ecosystem ensure that investors have access to accurate and timely information, mitigating the risk of fraud.

Ensuring Fair Governance (Governance Community): The Governance Community oversees the ecosystem's operations and maintains its integrity. They actively participate in decision-making processes and crack down on malicious activities, ensuring the interests of all parties are protected. This reduces the risk of conflicts of interest and ensures that the ecosystem's growth is sustainable and beneficial for all.

The payoff matrix for this three-player game can be represented as follows:

\begin{table}[ht]
\centering
\resizebox{\columnwidth}{!}{%
\begin{tabular}{|c|c|c|c|c|}
\hline
& $I_C$ & $N_C$ & $I_C$ & $N_C$ \\
\hline
$I_L, I_G$ & $(a, b, c)$ & $(d, e, f)$ & $(g, h, i)$ & $(j, k, l)$ \\
\hline
$N_L, I_G$ & $(m, n, o)$ & $(p, q, r)$ & $(s, t, u)$ & $(v, w, x)$ \\
\hline
$I_L, N_G$ & $(y, z, aa)$ & $(bb, cc, dd)$ & $(ee, ff, gg)$ & $(hh, ii, jj)$ \\
\hline
$N_L, N_G$ & $(kk, ll, mm)$ & $(nn, oo, pp)$ & $(qq, rr, ss)$ & $(tt, uu, vv)$ \\
\hline
\end{tabular}%
}
\end{table}

The tuples $(a, b, c)$, $(d, e, f)$, etc. represent the payoffs for the Labor Community, the Capital Community, and the Governance Community, contingent upon the strategies they adopt.

According to Nash's theorem, there exists a Nash equilibrium in this game, signifying a set of strategies $(I_C^, I_L^, I_G^*)$ such that no player can improve their payoff by unilaterally altering their strategy. This equilibrium can be represented by the following inequalities, implying that no player has an incentive to deviate from their chosen strategy, given the strategy of the other players.

For the Capital Community:
\begin{align*}
a \geq d, g, j \quad \text{if} \quad (I_L^{}, I_G^{}) = (I, I) \\
m \geq p, s, v \quad \text{if} \quad (I_L^{}, I_G^{}) = (N, I) \\
y \geq bb, ee, hh \quad \text{if} \quad (I_L^{}, I_G^{}) = (I, N) \\
kk \geq nn, qq, tt \quad \text{if} \quad (I_L^{}, I_G^{}) = (N, N)
\end{align*}

For the Labor Community and Governance Community, analogous inequalities can be inferred.

In this balanced interplay, all three communities benefit. The Capital Community sees a return on their investments, the Labor Community is rewarded for their contributions, and the Governance Community ensures the ecosystem's sustained growth and stability. This equilibrium promotes the equivalence between labor benefits and capital benefits, fostering a robust, interdependent ecosystem that embodies the principles of decentralized finance.

\subsubsection{Pareto Optimality in Ecosystem Communities}

Pareto optimality is a state of allocation of resources in which it is impossible to make any one individual better off without making at least one individual worse off. In other words, a situation is Pareto optimal when no individual can be made better off without making someone else worse off. In the context of the DCC ecosystem, we can consider a state to be Pareto optimal if no community (Capital, Labor, or Governance) can increase its benefits without reducing the benefits of at least one other community.

Let's consider the scenario where all communities choose to invest/contribute/govern (I). This is the state where:

The Capital Community invests on projects, stimulating activity and fostering the growth of the ecosystem.
The Labor Community creates valuable content and initiates crowdfunding projects, generating value within the ecosystem.
The Governance Community actively participates in decision-making processes and oversees project management, ensuring the integrity and fairness of the ecosystem.

In this state, all communities are benefiting. The Capital Community sees a return on their investments, the Labor Community is rewarded for their contributions, and the Governance Community ensures the ecosystem's sustained growth and stability.

Now, if any community decides to deviate from this strategy (I), it would disrupt the balance and negatively impact the ecosystem. For instance, if the Capital Community decides not to invest (N), it would limit the resources available for projects, thereby reducing the potential benefits for the Labor Community and the overall value generated within the ecosystem. Similarly, if the Labor Community decides not to contribute (N), it would lead to a lack of valuable content and projects, reducing the potential returns for the Capital Community and the overall vibrancy of the ecosystem. If the Governance Community decides not to govern (N), it could lead to unfair practices and instability in the ecosystem, negatively impacting all communities.

We could consider the allocation of resources X to the Capital Community (C), the Labor Community (L), and the Governance Community (G) as $(X_C, X_L, X_G)$. 

For the Capital Community:
{\small
\begin{align*}
U_C(X_C^*, X_L, X_G) \leq U_C(X_C, X_L, X_G) \quad \text{for all} \quad (X_C, X_L, X_G)
\end{align*}
}
For the Labor Community:
{\small
\begin{align*}
U_L(X_C, X_L^*, X_G) \leq U_L(X_C, X_L, X_G) \quad \text{for all} \quad (X_C, X_L, X_G)
\end{align*}
}
And for the Governance Community:
{\small
\begin{align*}
U_G(X_C, X_L, X_G^*) \leq U_G(X_C, X_L, X_G) \quad \text{for all} \quad (X_C, X_L, X_G)
\end{align*}
}
Where $U_C$, $U_L$, and $U_G$ are the utility functions of the respective communities.

Therefore, the state where all communities choose to invest/contribute/govern (I) is Pareto optimal in the DCC ecosystem. Any deviation from this state would make at least one community worse off, which aligns with the definition of Pareto optimality.

\subsection{Blockchain Consensus Mechanism Catalyzes Crowdfunding Enthusiasm through Benefit-sharing}

The innovation of blockchain technology has given rise to a diverse array of consensus mechanisms, each with their unique traits and implications for the performance and stability of the underlying blockchain ecosystem. In this section, we delve into the intricacies of various consensus mechanisms including PoW, PoS, and PBFT, their impact on the system processing capacity, and their ideological function within their respective public chain ecosystems.

The dynamics of Transaction Per Second (TPS) parameter, a key indicator of the system processing capacity of a consensus mechanism, provides vital insights into the performance characteristics of different blockchain consensus mechanisms \cite{wu2020hybrid}.

\begin{figure*}[htbp]
	\centering
	\begin{minipage}[b]{.49\linewidth}
		\includegraphics[width=\linewidth]{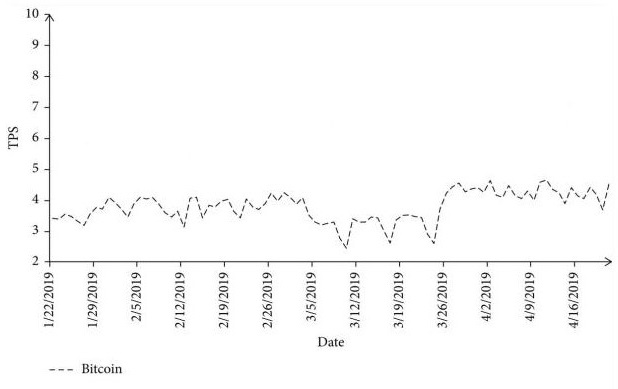}
		\caption{POW (BTC): TPS-TIME}\label{fig4-8-1}
	\end{minipage}
	\begin{minipage}[b]{.49\linewidth}
		\includegraphics[width=\textwidth]{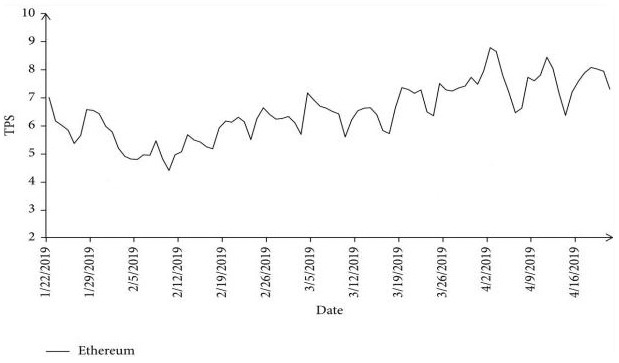}
		\caption{POS (ETH): TPS-TIME}\label{fig4-8-2}
	\end{minipage}
\end{figure*}

Figure \ref{fig4-8-1} illustrates that Bitcoin's Proof of Work (PoW) consensus mechanism, although exhibiting lower performance metrics, provides stable TPS values within the range of 2 to 4. This characteristic offers a predictable transaction processing capability despite its lower overall performance.

On the other hand, Ethereum employs the Proof of Stake (PoS) consensus mechanism as depicted in Figure \ref{fig4-8-2}, where TPS demonstrates greater variability and growth over time compared to the PoW mechanism. Despite this variability, the average performance of the PoS mechanism outperforms that of PoW.

\begin{figure*}[htbp]
	\centering
	\begin{minipage}[b]{.49\linewidth}
		\includegraphics[width=\textwidth]{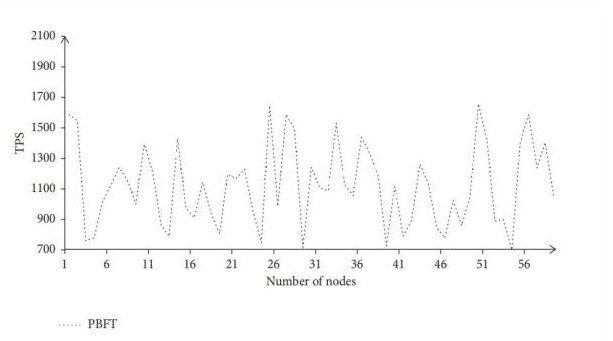}
		\caption{PBFT: TPS-Number of nodes }\label{fig4-8-3}
	\end{minipage}
	\begin{minipage}[b]{.49\linewidth}
		\includegraphics[width=\linewidth]{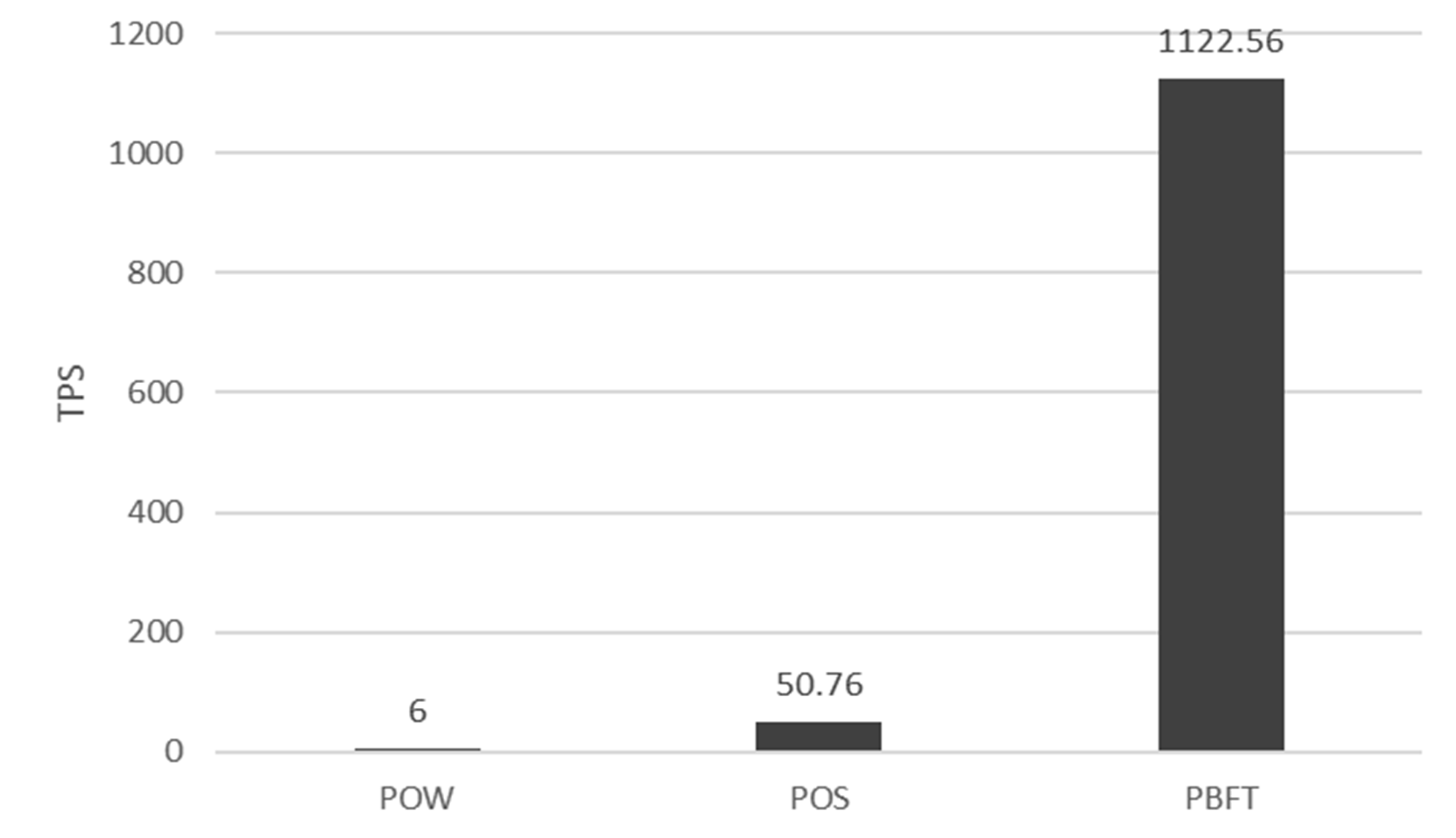}
		\caption{TPS under different consensus algorithms}\label{fig4-8-4}
	\end{minipage}
\end{figure*}

Investigation into the Byzantine Fault Tolerance mechanism, integral to the PBFT consensus process, shows a correlation between TPS changes and node count, as represented in Figure \ref{fig4-8-3}. PBFT's performance is significantly influenced by node diversity, and although it experiences instability in its performance coefficient, it surpasses PoW and PoS in handling multiple concurrent performance methodologies. An analysis of the TPS parameters of these mainstream consensus mechanisms (Figure \ref{fig4-8-4}) suggests a trade-off between system performance level and stability, and hints at a complementary relationship between the two.

\begin{figure*}[htbp]
	\centering
	\begin{minipage}[b]{.49\textwidth}
		\includegraphics[width=\textwidth]{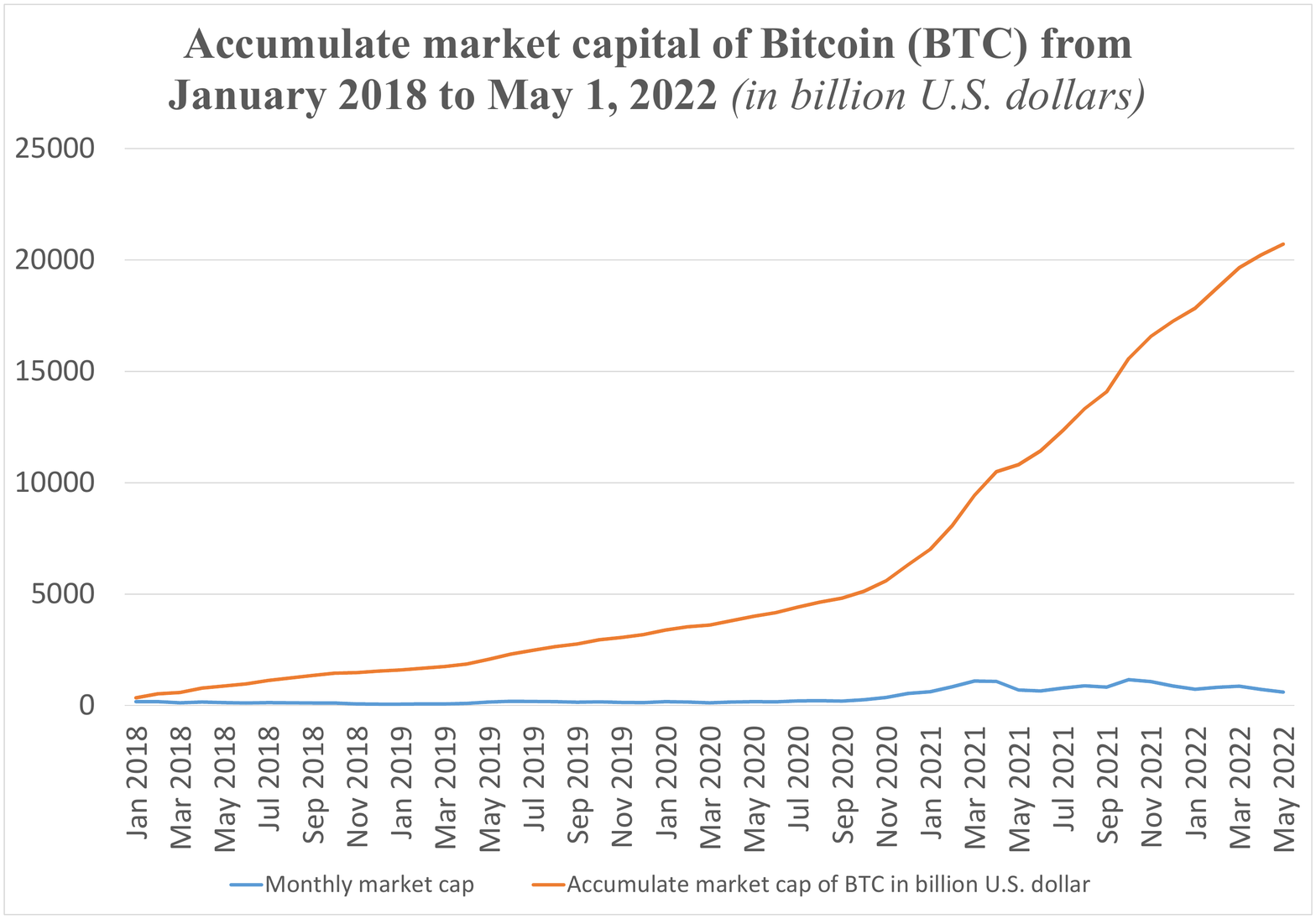}
		\caption{Accumulate market (ETH)}\label{fig4-8-5}
	\end{minipage}
	\begin{minipage}[b]{.49\textwidth}
		\includegraphics[width=\textwidth]{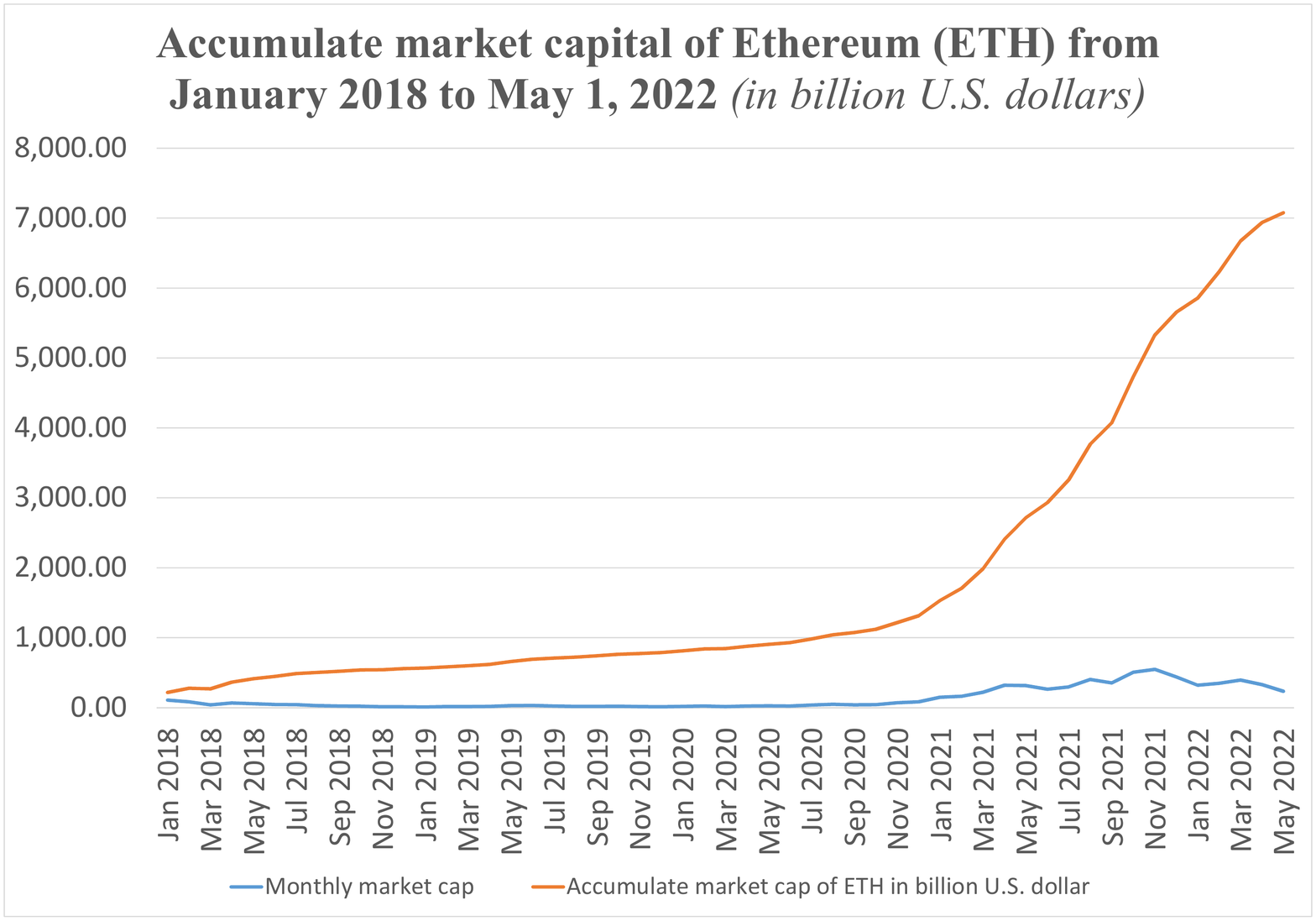}
		\caption{Accumulate market (BTC)}\label{fig4-8-6}
	\end{minipage}
\end{figure*}

The consensus mechanism not only serves as a technological tool, but also as a guiding ideology within public chain economic activities. The effectiveness and performance of the consensus mechanism directly impact the growth trajectory of the public chain economy. The growth curves of Bitcoin (PoW) and Ethereum (PoS) share similarities, as evident in Figures \ref{fig4-8-5} and \ref{fig4-8-6}. This resemblance underscores that both workload-based direct democracy (PoW) and organizational activity rights systems (PoS) are effective in attracting market interest.

\begin{figure*}[htbp]
	\centering
	\begin{minipage}[b]{.49\textwidth}
		\includegraphics[width=\textwidth]{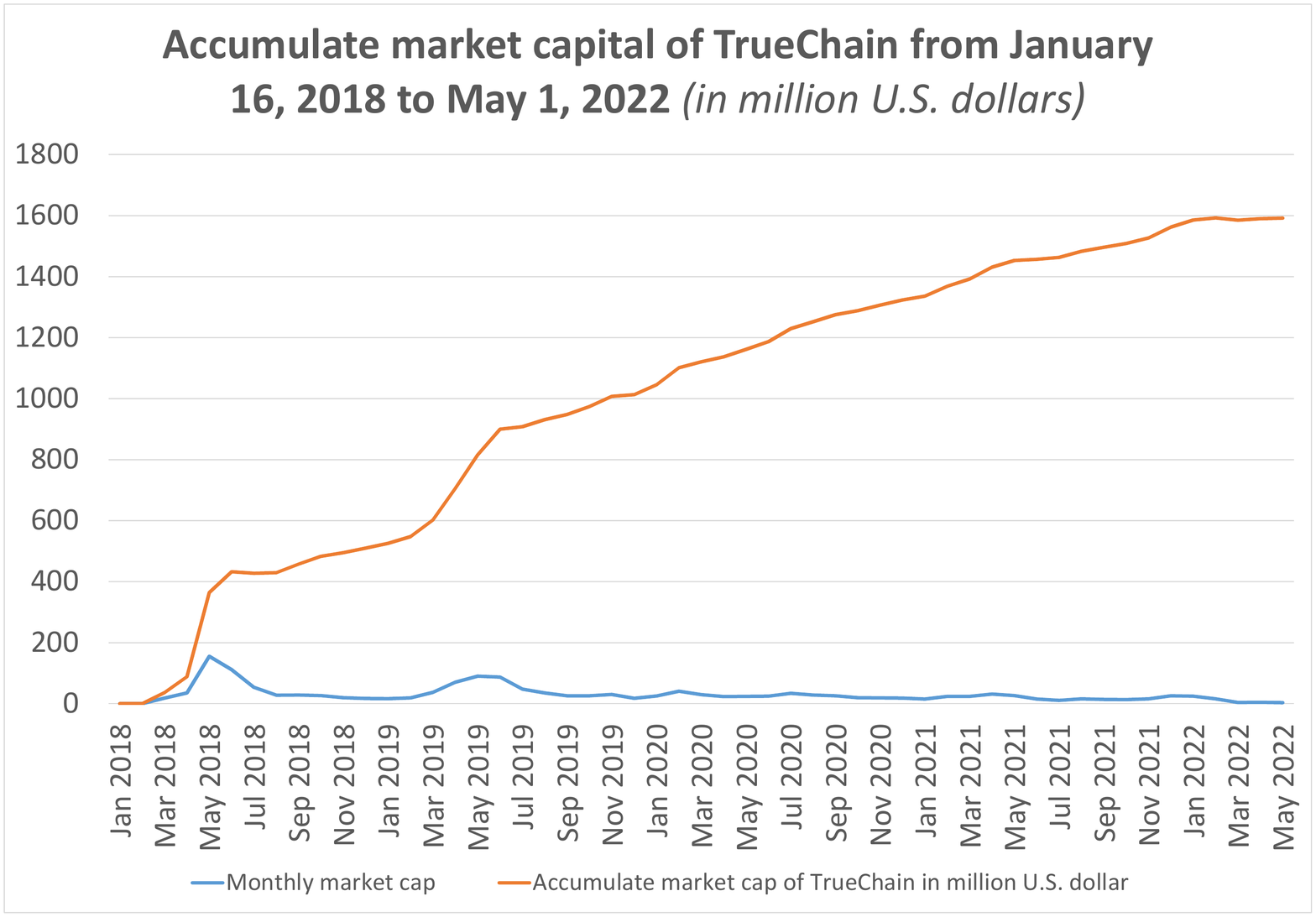}
		\caption{Accumulate market (Truechain)}\label{fig4-8-7}
	\end{minipage}
	\begin{minipage}[b]{.49\textwidth}
		\includegraphics[width=\textwidth]{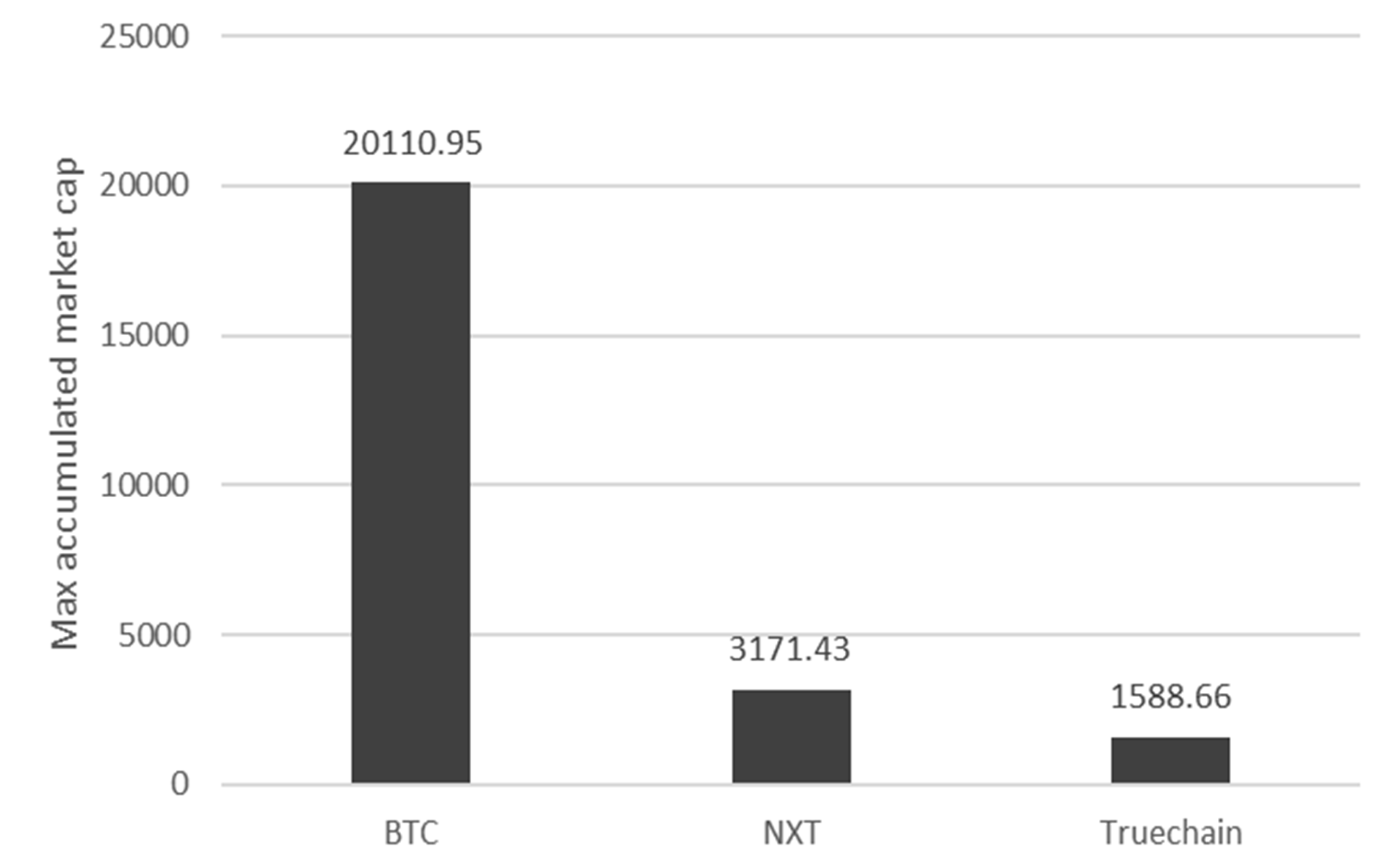}
		\caption{Max accumulate market cap under different consensus algorithms}\label{fig4-8-8}
	\end{minipage}
\end{figure*}

Moreover, Truechain, employing a PBFT consensus mechanism, presents a relatively stable market growth curve as portrayed in Figure \ref{fig4-8-7}. PBFT primarily targets ecosystems that form part of alliance chains and is designed to address the Byzantine Generals' problem. In this representational democracy model, organization managers form the majority of the audience groups. According to an analysis of cumulative market volume and audience growth in Figure \ref{fig4-8-8}, the consensus mechanism therefore ensures democratic rights at each community node, while also influencing the economy based on its underlying ideology.

Conventionally, the crowdfunding economy emphasizes shared benefits following shared risks. Sharing benefits encourages all groups within the economic framework to actively participate in the sharing movement. However, as this engagement is often driven by interests rather than trust, there is an inherent risk of economic activities being disproportionately influenced by dominant interest groups \cite{arrow1981risk}. Blockchain's consensus process aims to build trust before risk sharing, transforming weak trust into strong trust, and facilitating consensus to steer the sustainable development of the crowdfunding ecosystem. Consequently, our DPoLR consensus mechanism acts as a catalyst, fostering crowdfunding enthusiasm through shared benefits and providing a robust foundation for sustainable economic growth. This innovation ensures the sustainability of the crowdfunding ecosystem by fostering a democratic environment, wherein the influence of parties is dependent on the total Governance Tokens held, not capital, further enriching the decision-making environment. As such, the DPoLR mechanism effectively blends the benefits of stability, performance improvements, and inclusive governance to drive crowdfunding enthusiasm and sustainable economic growth.

\subsection{Economic Model of Crowdfunding Token Circulation and Valuation}

The economic worth of creative projects in the crowdfunding space can be ascertained using two fundamental metrics: Cash Flow, a direct measure of economic worth, and Total Information Quantity, representing the total token value circulation which signifies the economic value of crowdfunding \cite{magni2011aggregate}. In this ecosystem, Invalid Information is construed as the payback setting of the complete quantity of information, directly proportionate to the repayment demand of current liabilities to the project. As such, Invalid Information tends to reduce the token circulation ratio and the economic worth of the community.

Solow's research provides a foundation for the concept of Community Stability, which represents a balanced state where various components of the ecosystem continually create and retain value. In the domain of token economics, undeployed tokens held by the labor and capital community are seen as token deposits of individual creators, with tokens including both Labor Tokens and Capital Tokens. Looking at the bigger picture from a macroeconomic view, these total token deposits from the labor and capital community act as a benchmark for projecting the value of future token circulation. This benchmark mirrors the result one would get from a compound interest present value calculation.

Tokens play a pivotal role in the crowdfunding token economy, serving as a universal unit in the commodity exchange process. They form a vital link connecting the fundraising concept to the capital provided by funders and the endorsement given by curators, represented by invested Capital Tokens and pledged Labor Tokens respectively. Consequently, the principles of the monetary banking formula and the money supply formula govern the availability and liquidity of tokens in this system.

As shown in Table \ref{table4-1}, within the blockchain-based crowdfunding economy, the prices of crowdfunding projects are determined by the number of absorbed tokens. This number, in combination with token value, dictates the price of the crowdfunding projects. Funders essentially purchase commodities based on the price of the fundraising project, which is set in terms of the amount of absorbed tokens. Evaluators then use Labor Tokens to assess the value of the project.

\begin{table*}[htbp]
	\centering
	\includegraphics[width=1\textwidth]{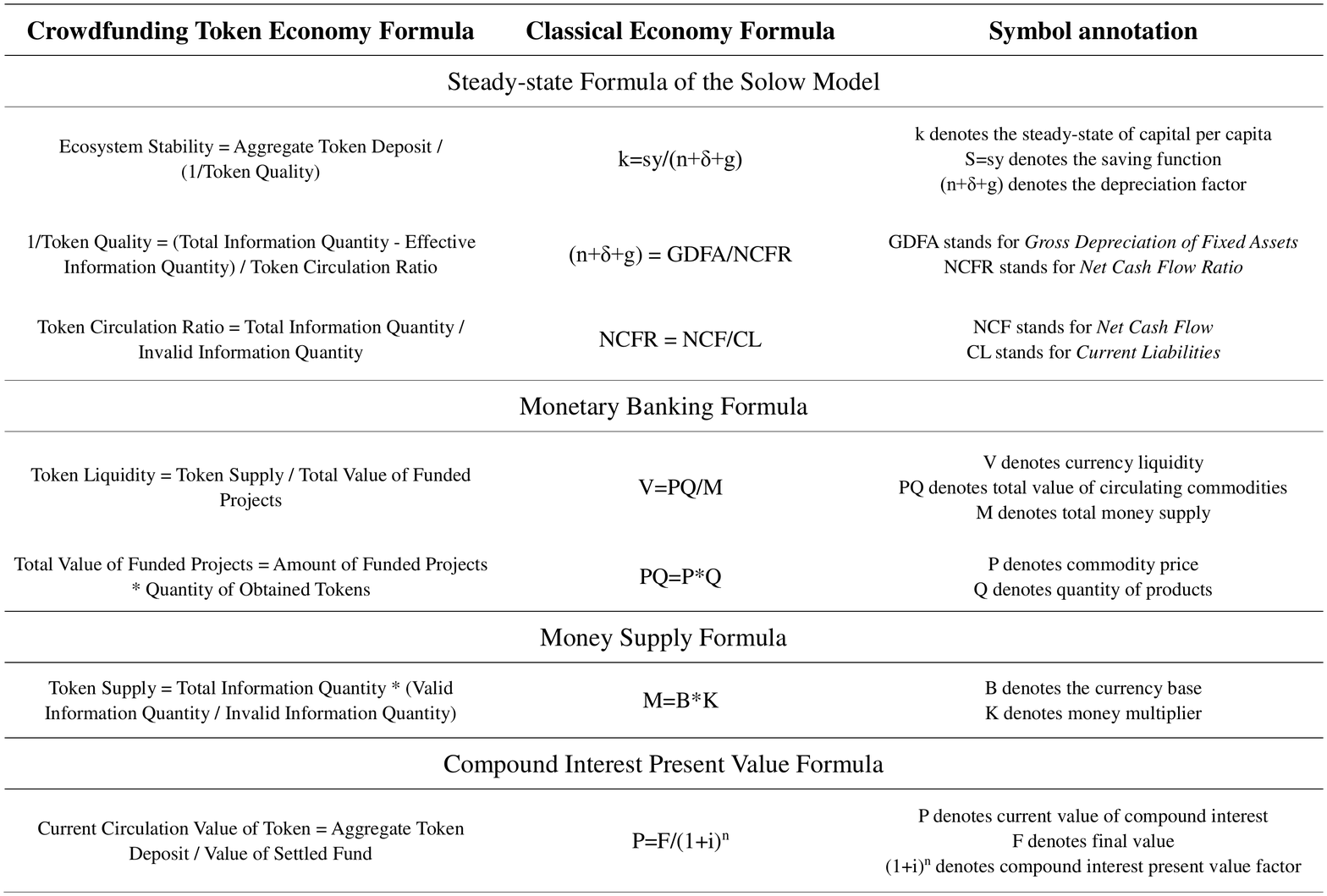}
	\caption{The comparison of formulas between crowdfunding token economy and classical economy}\label{table4-1}
\end{table*}

\subsection{Adjustments in the Crowdfunding Token Economy to Stabilize the Ecosystem}

In a crowdfunding token economy, the production of tokens is a shared responsibility between the labor community and the governance community. This process of token generation falls under the umbrella of production supervision within the blockchain-based crowdfunding economy system, which essentially entails reviewing the content of on-chain data. The governance community plays a crucial role in monitoring and controlling this process. While qualified commodities find their way into the market as supply commodities, production supervision exerts a positive influence on the quality of commodities and inversely affects the quantity of commodities. Table \ref{table4-1} illustrates the comparison of formulas between the crowdfunding token economy and classical economy.

However, an imbalance can occur during the distribution phase, notably if there is a substantial bias towards the governance community. Such bias may cause a decrease in interest towards information production and its associated regulatory activities, which intentional governance nodes can engage in. This could lead to a reduction in the quantity of labor distribution tokens that can be held, adversely affecting the interests of both governance nodes and creator nodes. Therefore, the governance community has the capacity to stimulate production by tweaking the interest distribution when the production of valid information is in decline.

Two distinct scenarios could lead to an increase in both the token circulation ratio and the token inflation ratio. The first scenario, according to monetary quantity theory, unfolds when there is an increase in token supply and total information quantity while supervision strength remains constant. This results in an increase in the number of absorbed tokens. However, as the average supervision intensity drops, both valid information quantity and token quality will decrease. This disrupts the equilibrium of ecosystem stability and its steady state. Consequently, tokens fail to meet the benchmark that is foundational for fundraisers' evaluations. To counteract this decline in token quality, governance groups can decrease the production willingness of inaccurate information producers by reducing the labor token distribution quota through consensus adjustments. Alternatively, they can enhance effective supervision and token quality by increasing supervision incentives.

The second scenario happens when the production oversight decreases while the token supply and total information quantity remain unchanged. This leads to a reduction in valid information quantity and, subsequently, a decline in token quality. By optimizing the incentive allocation for effective oversight, the governance group can elevate token quality, thus ensuring the stability of the ecosystem.

\subsection{Decentralized Co-governance Towards an Alternative Communist Social Pattern}

The Decentralized Co-governance Crowdfunding (DCC) Ecosystem embodies an intriguing trajectory that parallels key stages of economic theory, albeit with unique adaptations. It unfolds a narrative spanning the capitalist, socialist, and communist modes of socio-economic interaction, mirroring an alternative communist social pattern within the ecosystem.

In the early stages of the ecosystem, the dynamics mirror a capitalist system. During this transitional phase, the labor value is underestimated, and the exchange rate between Labor Tokens and Capital Tokens is below the equilibrium level. This discrepancy allows the Capital Community to gain disproportionately through investments, capital operations, and financial services, reaping benefits that surpass their contribution to liquidity promotion and resource allocation. The Capital Community, hence, temporarily obtains surplus income, exploiting the Labor Community and stripping them of surplus value. However, this phase is temporary due to the increase of Labor Tokens' value, driven by the tokens' scarcity and inaccessibility.

As the ecosystem matures, it transitions into a socialist stage. During this quasi-steady state, the value of labor reaches a reasonable level, aligning the gains of the Capital Community closely with the value they contribute to the ecosystem. This equilibrium incites a negative feedback regulation mechanism: when Labor Tokens value drops below the steady-state level, labor participation becomes less rewarding. The decrease in labor supply leads to an increase in the value of labor tokens. On the contrary, when Labor Tokens value inflates, labor participation becomes more rewarding, leading to reduced ecosystem liquidity and excess supply of Labor Tokens, and consequently, a decrease in the value of Labor Tokens. This equilibrium makes it difficult for the Capital Community to exploit the surplus value of the Labor Community.

Ultimately, the ecosystem evolves into its final stage, resembling a communist state. This evolution is fueled by users' improved ideological awareness and moral standards, resulting in fewer invalid information generation, decreased governance costs, and eventual dissolution of the Governance Community. This state paves the way for substantial productivity improvements, taking the ecosystem to a highly developed level owing to the minimized cost of governance.

In this communist theory, several unique principles underpin the ecosystem's functioning. Distribution is aligned with labor, with Labor Tokens epitomizing the principle of "distribution according to work". The ecosystem also ensures relative fairness as the disparity in labor value generated by different individuals is less pronounced than the existing wealth gap. The scarcity and hard availability of Labor Tokens embed the idea that "labor is the most glorious", and the prohibition of Capital Tokens in governance encapsulates the concept of 'laborers as the masters'.

This unique approach to communism is shaped by the DCC ecosystem. The system reframes existing class divisions based on token ownership, allowing individuals to belong to multiple classes simultaneously. Here, class merely signifies corresponding rights and responsibilities, devoid of any hierarchical distinction or political connotations, hence eliminating antagonism.

In the DCC ecosystem, conventional class divisions like proletariat, petty bourgeoisie, middle class, bourgeoisie, etc. are replaced with the labor class and capital class. This approach aligns with the blurred class boundaries in modern societies and acknowledges that laborers may not necessarily be proletarian. Furthermore, the ecosystem incorporates elements of modern economics, considering management costs, transaction costs, financial costs, etc., and updating the surplus value theory. It recognizes the necessity of the capital class and a market economy within a socialist and even communist society. The ecosystem achieves common ownership not through eliminating or confiscating private property, but through token stakes held by all users.

The DCC ecosystem innovatively proposes that socialism enhances labor value and labor income through market self-adjustment and price mechanism, effectively curtailing the capital class's exploitation of surplus value. This dynamic propels the capital class to support the labor class in generating greater value, ultimately benefitting both communities. Nevertheless, under the DCC ecosystem's distribution mechanism, labor receives a relatively larger share, promoting a balance between labor and capital benefits. As the ecosystem evolves and its overall value expands, this proposed distribution mechanism ensures a fair and more equitable sharing of benefits. This system allows for the transition to socialism and communism to occur organically, without necessitating fierce struggle. As such, the DCC ecosystem offers a novel alternative to classical communism, utilizing digital tokens and decentralized governance mechanisms to engender a more equitable distribution of value and wealth.

\section{Conclusion}

In this paper, we have meticulously explored the Decentralized Co-governance Crowdfunding (DCC) Ecosystem, a novel approach that not only addresses the inherent issues in conventional crowdfunding platforms but also introduces a new paradigm in the realm of blockchain crowdfunding. This ecosystem, with its unique interplay between Capital, Labor, and Governance communities, promotes sustainability, transparency, mutual trust, safety, and supervision in a sociological and economic landscape.

The DCC model redefines how we view labor, capital, and governance in token economies, bringing forth a new perspective that embodies democratic, equitable, and just principles. These profound implications for research and practice open avenues for future exploration and study, while also resolving the challenges faced by conventional crowdfunding platforms.

\subsection{Implications to Research and Practice}

The DCC ecosystem provides a practical embodiment of theoretical concepts related to labor, capital, and governance in a decentralized digital economy. It offers a unique perspective on blockchain-based socioeconomic systems, redefining the roles of labor and capital, and shedding new light on blockchain-based democratic governance mechanisms.

For practitioners, especially those involved in the burgeoning field of DeFi, the DCC ecosystem presents a robust model for developing and managing decentralized platforms that balance labor value, capital investment, and governance responsibility. It promotes transparency, mutual trust, safety, and supervision, suggesting ways to build and sustain a token economy where all participants have an opportunity to contribute, benefit, and influence decision-making processes. This realization could significantly impact how future digital economies function, fostering more equitable, efficient, democratic, and just financial systems.

\subsection{Limitations and Suggestions for Future Research}

While this paper has mainly focused on the qualitative understanding of the DCC ecosystem, the lack of quantitative analysis is a limitation. The evaluation of the DCC model largely remains theoretical, and practical implications may unveil challenges not yet anticipated. The proposed transition from capitalism to an alternative form of communism through a token-based economy necessitates further exploration and validation. The applicability of this framework in real-world contexts, especially under varying cultural, regulatory, and economic conditions, remains untested.

Future studies should emphasize quantitative research and empirical testing in sociology and economics to substantiate our claims about the DCC ecosystem. This could include empirical studies, simulations, or econometric modeling. Furthermore, a comparative analysis of the proposed evolutionary socio-economic model with existing ones could help gauge its potential impact on wealth distribution, social equity, economic development, and the promotion of transparency, mutual trust, safety, and supervision.









\bibliographystyle{unsrt}

\end{document}